\documentclass[twocolumn,showpacs,preprintnumbers,amsmath,amssymb,aps,pra,superscriptaddress,10pt]{revtex4-1}

\usepackage[latin1]{inputenc}
\usepackage{dcolumn,color}
\usepackage{bm}
\usepackage{graphicx}
\usepackage[english]{babel}

\def\bbm[#1]{\mbox{\boldmath $#1$}}
\newcommand{\atwo}{a_{2\rm D}^{\rm eff}}
\newcommand{\rr}{{\bf r}}
\newcommand{\rrho}{{\boldsymbol\rho}}
\newcommand{\GGamma}{{\boldsymbol\Gamma}}
\newcommand{\aaa}{{\bf a}}
\newcommand{\bbb}{{\bf b}}
\newcommand{\qq}{{\bf q}}
\newcommand{\pp}{{\bf p}}
\newcommand{\RR}{{\bf R}}
\newcommand{\KK}{{\bf K}}
\newcommand{\ttt}{{\bf t}}
\newcommand{\HH}{\mathcal{H}}

\newcommand{\PV}{\mathcal{P}}
\newcommand{\MM}{\mathbb{M}}
\newcommand{\II}{\mathbb{I}}
\newcommand{\TT}{\mathbb{T}}
\newcommand{\AAA}{\mathcal{A}}

\begin{document}

\title{Matter waves in two-dimensional arbitrary atomic crystals}

\author{Nicola Bartolo}\email{nicola.bartolo@univ-montp2.fr}
\affiliation{Universit\'{e} Montpellier 2, Laboratoire Charles Coulomb UMR 5221 - F-34095 Montpellier, France}
\affiliation{CNRS, Laboratoire Charles Coulomb UMR 5221 - F-34095 Montpellier, France}
\affiliation{INO-CNR BEC Center and Dipartimento di Fisica, Universit\`{a} di Trento - I-38123 Povo, Italy}
\author{Mauro Antezza}\email{mauro.antezza@univ-montp2.fr}
\affiliation{Universit\'{e} Montpellier 2, Laboratoire Charles Coulomb UMR 5221 - F-34095 Montpellier, France}
\affiliation{CNRS, Laboratoire Charles Coulomb UMR 5221 - F-34095 Montpellier, France}
\affiliation{Institut Universitaire de France - 103, bd Saint-Michel - F-75005 Paris, France}

\date{\today}

\begin{abstract}
We present a general scheme to realize a cold-atom quantum simulator of bi-dimensional atomic crystals.
Our model is based on the use of two independently trapped atomic species: a first one, subject to a strong in-plane confinement, constitutes a 2D matter wave which interacts only with atoms of a second species, deeply trapped around the nodes of a 2D optical lattice.
By introducing a general analytic approach we show that the system Green function can be exactly determined, allowing  for the investigation of the matter-wave transport properties. We propose some illustrative applications to both Bravais (square, triangular) and non-Bravais (graphene, kagom\'e) lattices, studying both ideal periodic systems and experimental-size and disordered ones.
Some remarkable spectral properties of these atomic artificial lattices are pointed out, such as the emergence of single and multiple gaps, flat bands and Dirac cones. All these features can be manipulated via the inter-species interaction, which results widely tunable due to the interplay between scattering length and confinements.
\end{abstract}

\pacs{03.75.-b, 37.10.Jk, 67.10.Jn, 67.85.-d}

\maketitle

\section{Introduction}

In spite of the development of supercomputers and cutting-edge numerical methods, the simulation of experimental size  many-body systems is still a hard task.
Following Feynman's conjecture of a quantum simulator, it results useful to analyze manipulable systems able to reproduce, in a controllable way, the physical scenario that one wants to investigate.
The theoretical and experimental progresses of the last decades, boosted by nano-technological needs, made possible to combine the properties of cold gases and optical lattices to build-up artificial crystals able to mimic condensed-matter systems \cite{BlochNature12}.
A crucial role in the success of these models is due to their experimental versatility.
Interactions between cold atoms can indeed be tuned via Feshbach \cite{ChinRMP10}, dipolar \cite{CitDIR} or confinement-induced \cite{OlshaniiPRL98} resonances. On the other hand optical lattices offer a full control of the potential landscape felt by cold atoms, allowing for the exploration of quantum phase transitions \cite{CitQPT}.
Furthermore such kind of simulators enable to explore parameter ranges beyond those of the real material they imitate, unveiling new physical scenarios.

\begin{figure}[tb]
\includegraphics[width=0.45\textwidth]{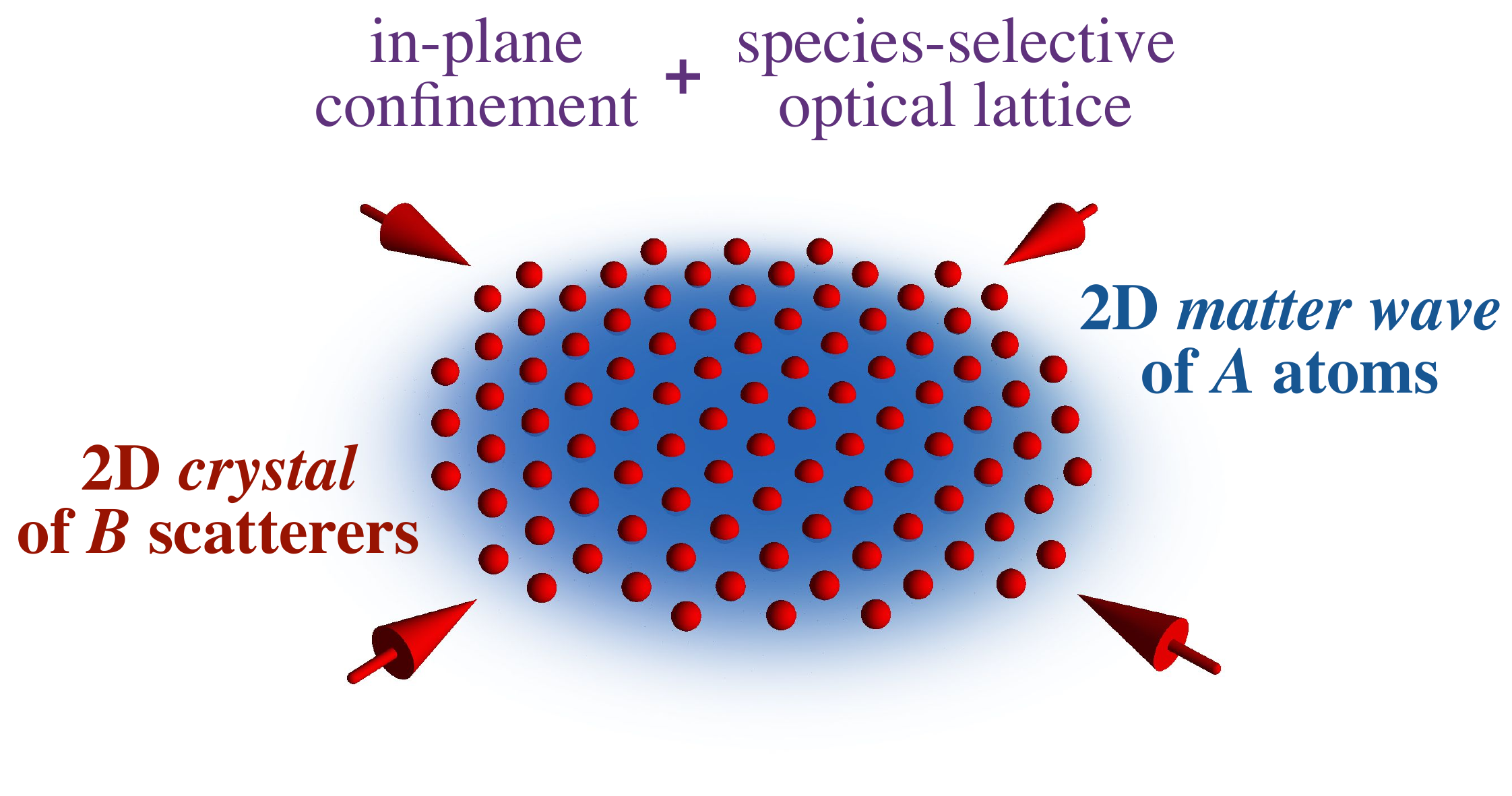}
\caption{(Color online)
Schematic representation of our model for the realization of two-dimensional atomic crystals.
Two atomic species, namely $A$ (depicted in blue) and $B$ (depicted in red) are strongly confined on a plane.
Making use of a species-selective optical lattice \cite{LamporesiPRL10,GadwayPRL11} $B$ atoms are arranged in a 2D lattice of point-like scatterers of arbitrary geometry (a square one in this example) while $A$ atoms form a matter wave which propagates through the artificial crystal.}
\label{FigSystem}\end{figure}

In solid state physics, among the plethora of crystals that can be investigated, two-dimensional ones are of special interest due to the intriguing properties that 2D materials have shown to posses.
Up to the early 2000s the study of these systems was only of academic interest, 2D solids being considered unstable structures never observed experimentally.
Things changed in 2004, when graphene was finally isolated \cite{NovoselovScience04}. This discover paved the way to the study of this astonishing carbon allotrope constituted by a mono-layer of ions forming an honeycomb lattice in which charge carriers manifest peculiar transport properties \cite{GeimNature07,CastroNetoRMP09}.
In particular conduction and valence bands touch in isolated points of $k$-space: the Dirac points. Around them the energy-momentum dispersion relation is conical and a Dirac-like equation for massless fermions replaces Schr\"odinger's one to describe the quantum motion of the carriers.
Graphene qualifies thus as a quantum electrodynamics simulator on a benchtop scale.
Furthermore relativistic effects, in general inversely proportional to the speed of light, would be enhanced in graphene: the role of $c$ is here played by the group velocity $v_g$ of the particles around the cone and $c/v_g\!\sim\!300$ \cite{GeimNature07}.

The growing attention for graphene and others mono-layer materials translates in an increasing interest toward their quantum simulators, so that many artificial prototypes of 2D lattices have been proposed and realized in the past years (for a recent review see \cite{PoliniNature13}).
In the present work we introduce a general, highly controllable model for the realization of artificial bi-dimensional lattices, based on the use of two cold-atomic species.
In our system a 2D matter wave, made of $A$ atoms, interacts with point-like scatterers of a second atomic species, denoted by $B$, independently trapped around the nodes of a 2D optical lattice.
A schematic representation of our model is presented in Fig.~\ref{FigSystem}.
Such a scheme is already experimentally realizable using species-selective optical lattices: trapping potentials engineered to act on an atomic species ($B$ in present case) being at the same moment invisible to a second one (for us $A$) \cite{LamporesiPRL10,GadwayPRL11}.
This has been done, for instance, in \cite{LamporesiPRL10} for a mixture of $^{87}$Rb and $^{41}$K atoms: tuning the optical-lattice frequency exactly in between two $^{87}$Rb resonances the attractive and repulsive contribution to the optical force cancel each other and only $^{41}$K feels the added potential.
In this study we limit our investigations to one-body physics in the matter wave, i.e. we assume $A$ atoms to be non-interacting with themselves, situation attainable by using polarized fermions or bosons at zero scattering length.
One can instead employ the $B\!-\!B$ interaction to reach a Mott insulating phase with exactly one atom per lattice site \cite{CitQPT} and subsequently freeze the atoms in this configuration by increasing the lattice depth. Other techniques are also available to probe \cite{BakrScience10} and manipulate \cite{NogrettePRX14} at single-site and single-atom level the scatterers arrangement.
This model has been recently proposed to study the effects of disorder in 1D \cite{GavishPRL05}, 2D and 3D systems \cite{AntezzaPRA10}, and it has the advantage to show a one-to-one correspondence with the bi-dimensional lattices that can mimic: the $A$ atoms of the matter wave play the role of the electronic cloud while the deeply trapped $B$ ones represent the crystalline structure.
From now on we will refer to our system as atomic artificial crystal (AAC), since the periodic potential felt by the matter wave is generated by other atoms and not by a sub-standing optical potential.
 
The paper is organized as follows.
We start by introducing the theoretical model in Sec.~\ref{SecTheoreticalModel}, briefly discussing the problem of scattering in reduced \emph{and} mixed dimensions (Sec.~\ref{SecScattering}) and describing the general approach to the study of AAC in Sec.~\ref{SecGeneral}.
In Sec.~\ref{SecBravais} we specify the model to the case of ideal Bravais arrangements of the scatterers, studying as illustrative examples the square (Sec.~\ref{SecSquare}) and triangular (Sec.~\ref{SecTriangular}) lattices. For both of them the spectral properties are analyzed in function of the $A\!-\!B$ interaction strength. Finite-size and disorder effects are also investigated.
Sec.~\ref{SecNonBravais} is devoted to the generalization of previous results to non-Bravais lattices, focusing on the exemplary atomic artificial graphene (Sec.~\ref{SecGraphene}) \cite{BartoloEPL14} and kagom\'e lattice (Sec.~\ref{SecKagome}). Some properties of these systems, namely the emergence of Dirac cones and non-dispersive flat bands, are pointed out and characterized.
We finally present our conclusions in Sec.~\ref{SecConclusions}.

\section{Theoretical model}\label{SecTheoreticalModel}

\subsection{0D-2D scattering process}\label{SecScattering}

A remarkable feature of cold-atomic systems is the possibility to use experimentally controllable parameters as knobs to tune the interatomic scattering lengths, for example by means of Feshbach or dipolar-induced resonances \cite{ChinRMP10,CitDIR}.
When the interacting particles are subject to a trapping potential, this can, in turn, play a role leading to confinement-induced resonances \cite{OlshaniiPRL98}. If the trapping is sufficiently strong the dimensionality of the system can be reduced: it has been proved, for instance, that a 3D system subject to a strong axial confinement (quasi-2D) can be mapped into a strictly-2D one by introducing an effective 2D scattering length, the latter depending on the free 3D interaction and on the trapping parameters \cite{PetrovPRA01}.
Another related subject is the study of scattering processes involving differently-trapped atoms, in this cases we talk about scattering in mixed dimensions. Recently processes involving a free particle and a trapped one have been addressed (nD-3D scattering \cite{MassignanPRA06,NishidaPRL08PRA10}) and some theoretical predictions have been tested experimentally \cite{LamporesiPRL10}.

\begin{figure}[tb]
\includegraphics[width=0.45\textwidth]{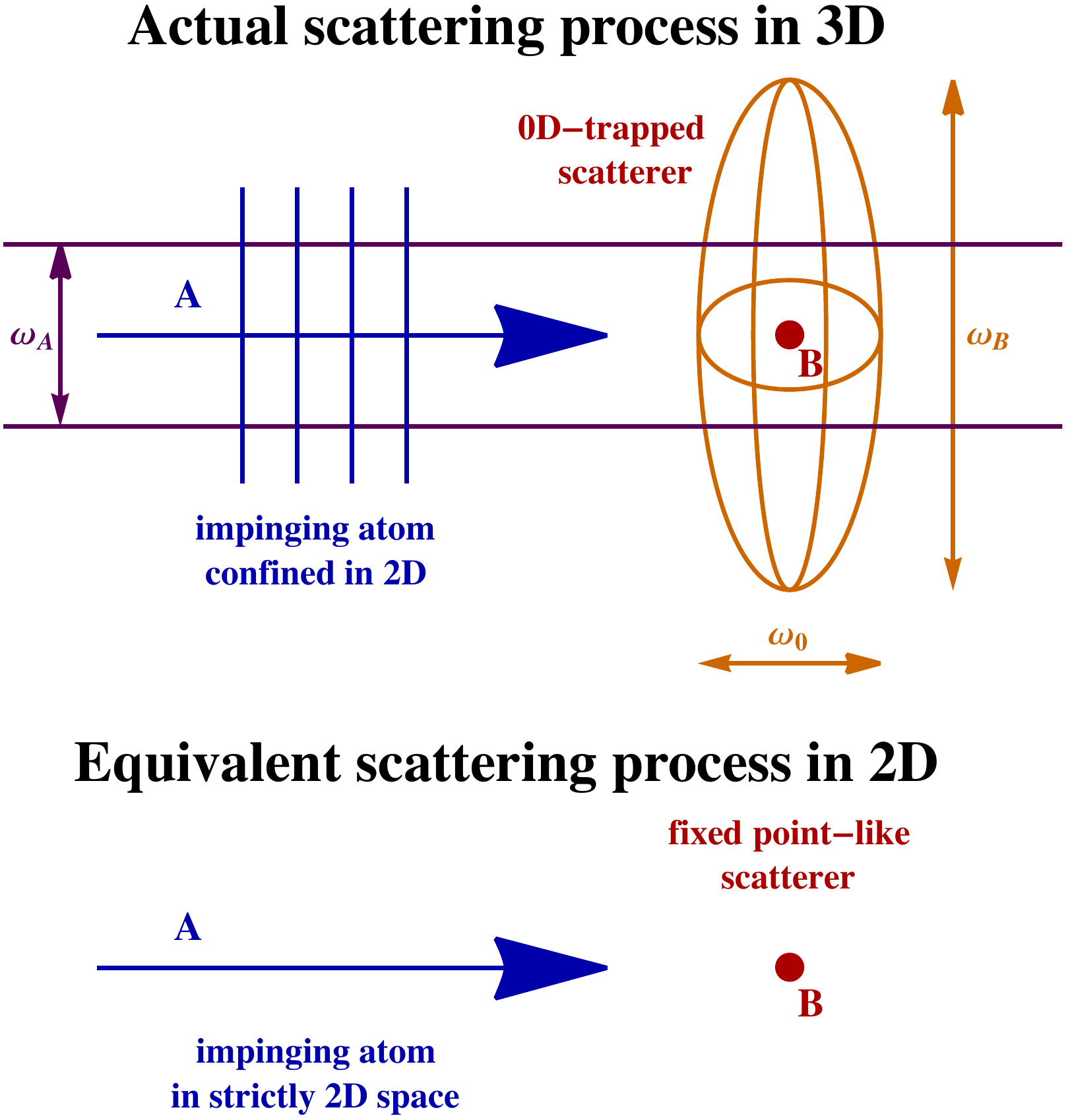}
\caption{(Color online)
Schematic representation of the 2-body scattering process between a 2D-trapped $A$ atom and a 0D-trapped $B$ one.
The scattering process in the actual system (top panel) is ruled by the 3D $A\!-\!B$ scattering length $a_{3\rm D}$ of the contact interaction, as well as by the trapping frequencies and the mass ratio of the two atomic species.
The system is equivalent to a strictly bi-dimensional one (bottom panel) in which the $B$ atom constitutes a fixed point-like scatterer and the scattering process is ruled only by an effective 2D scattering length $\atwo$ \cite{BartoloPreparation}.}
\label{FigScattering}\end{figure}

In the case of the AAC the building-block of any analysis is the 2-body low-energy scattering process between an $A$ atom of mass $m_{\!A}\!=\!m$, harmonically trapped on a plane, and a $B$ one of mass $m_{\!B}$, trapped around a node of an optical lattice, process schematically represented in Fig.~\ref{FigScattering}. It can be shown that, by providing a proper effective 2D scattering length $\atwo$, the system is mapped into a strictly 2D space in which the $B$ atom is now a fixed point-like scatterer. The effective parameter $\atwo$, which takes also into account the quantum motion of the $B$ atom in the real system, depends on the atomic mass ratio, the trapping frequencies and the $A\!-\!B$ scattering length $a_{3\rm D}$ in free space (c.f. Fig.~\ref{FigScattering}).
A detailed study of this 0D-2D scattering process will be extensively presented elsewhere \cite{BartoloPreparation}. For the purposes of present work we assume $\atwo$ to be tunable in its full range of existence $[0,\infty[$, protecting the validity of the point-like approximation for $B$.
In this regime the interaction can be taken into account by considering the $A$ atom as free and imposing the Bethe-Peierls contact conditions
\begin{equation}\label{BPCondition}
\psi(\rr)\xrightarrow[]{\rr\to\rr_{\!B}}
\frac{m}{\pi\hbar^2} D_{\!B} \ln\left(\frac{|\rr-\rr_{\!B}|}{\atwo}\right)
+O\left(|\rr-\rr_{\!B}|\right)
\end{equation}
on its wave function at the position $\rr_{\!B}$ of the scatterer, where $D_{\!B}$ is an arbitrary complex coefficient.
We finally recall that, in the case of 2D scattering, the limit $\atwo\!\to\!\infty$ corresponds to a weakly attractive interaction, while for $\atwo\!\to\!0$ the effective potential allows a single, infinitely deep bound state and results weakly repulsive for positive-energy scattering states \cite{OlshaniiPRL01JPB07}.

\subsection{General approach}\label{SecGeneral}

Let us start by considering now a general case in which $N$ point-like $B$ scatterers are fixed at positions $\{\rr_i\}$.
The first steps of our calculation follow \cite{AntezzaPRA10}: we consider the hamiltonian of the matter wave (MW) as that of a free $A$ atom, i.e. $\HH\!=\!-\frac{\hbar^2}{2m}\nabla_{\!2\rm D}^2$ with $\nabla_{\!2\rm D}$ the 2D Laplace operator, and add the $A\!-\!B$ interaction by imposing the boundary conditions \eqref{BPCondition} at the position of each $B$ scatterer. This introduces a set of $N$ independent complex coefficient $D_i$.
The same conditions apply to the MW Green's function $G(\rr,\rr_0)$, solution of the Schr\"odinger equation for a point-like source term of matter waves in $\rr_0$: $(E+i0^+\!-\HH)G(\rr,\rr_0)\!=\!\delta(\rr-\rr_0)$. The latter wave equation can be rewritten to take directly into account the boundary conditions. By using the identity $\nabla_{\!2\rm D}\ln(r)\!=\!2\pi\delta(\rr)$, the effect of contact conditions resumes in the inclusion of secondary point-like sources of amplitude $D_i$ at the position of each scatterer \cite{AntezzaPRA10}, leading to
\begin{equation}\label{GreenFunctionWaveEquation}
(E+i0^+\!-\HH)\,G(\rr,\rr_0)=
\delta(\rr-\rr_0)+\sum_{i=1}^N D_i\,\delta(\rr-\rr_i).
\end{equation}
Since the poles of $G$ (and of its analytical continuation to complex energies in the lower half-plane) correspond to eigenstates of the system, its knowledge is of fundamental importance to determine the properties of the AAC.

To integrate Eq.~\eqref{GreenFunctionWaveEquation} we use its solution in absence of scatterers, that is the case of a free 2D MW. In this case $G(\rr,\rr_0)\!=\!g(\rr-\rr_0)$ with
\begin{equation}\label{Defg0}
g_0(\rr)=-i\frac{m}{2\hbar^2}H_0^{(1)}(kr),
\end{equation}
where $H_0^{(1)}$ is the zero-index Hankel function of the $1^{\rm st}$ kind. The wave vector modulus $k$ is linked to the MW energy by $E\!=\!\hbar^2k^2\!/2m$, with $k\!>\!0$  for $E\!>\!0$ and $k\!=\!i\kappa$ with $\kappa\!>\!0$ for $E\!<\!0$ (i.e. for bound states).
The formal solution of the wave equation is hence
\begin{equation}\label{GreenFunctionWaveFunction}
G(\rr,\rr_0)=g_0(\rr-\rr_0)+\sum_{i=1}^N D_i\,g_0(\rr-\rr_i),
\end{equation}
where the determination of the $N$ coefficients $D_i$ depends on the system geometry, encoded in the set $\{\rr_i\}$.
This problem can be led back to the solution of a complex linear system in the $N$ unknowns $D_i$. Each equation of such system comes from the limit $\rr\!\to\!\rr_j$ of Eq.~\eqref{GreenFunctionWaveFunction}, imposing the Bethe-Peierls condition on the LHS and applying
\begin{equation}
H_0^{(1)}(kr)\xrightarrow[r\to0]{}
1+\frac{2i}{\pi}\ln\left(\frac{e^\gamma}{2}kr\right)+o(1)
\end{equation}
on the RHS, where $\gamma\!\simeq\!0.577216$ is the Euler-Mascheroni constant.
After some straightforward algebraic manipulation the system can be cast as
\begin{equation}\label{LinearSystem}
\sum_{i=1}^N \MM_{ji}D_i = -\frac{\pi\hbar^2}{m} g_0(\rr_j-\rr_0) \qquad j=1,2,\cdots,N,
\end{equation}
with the introduction of the matrix $\MM$ of elements
\begin{equation}\label{DefinitionM}
\MM_{ji}=
\begin{cases}
\frac{\pi\hbar^2}{m} g_0(\rr_j-\rr_i) &\rr_j\neq\rr_i\\
\ln\left(\frac{e^\gamma}{2}k\atwo\right)-i\frac{\pi}{2} & \rr_j=\rr_i.
\end{cases}
\end{equation}
The formal solution of $G$ (Eq.~\eqref{GreenFunctionWaveFunction}) has a pole if $\MM$ is not invertible, i.e. if $\det(\MM)\!=\!0$.
The latter is thus our general condition for the existence of an eigenstate of the MW in the gas of scatterers.

The condition $\det(\MM)\!=\!0$ can be rewritten in a more practical form by noticing that the interaction-dependent terms appear only in the diagonal elements of $\MM$. In particular one can write
\begin{equation}
\MM_{jj}=	\ln\left(\frac{e^\gamma}{2}ka\right)-i\frac{\pi}{2}+\alpha
\end{equation}
with the introduction of the 2D interaction coefficient $\alpha\!=\!\ln(\atwo/a)$ and for an arbitrary choice of the unitary length $a$. It follows that $\MM\!=\!\MM^o+\II\alpha$, for $\MM^o\!=\!\MM(\alpha\!=\!0)$ and $\II$ the $N\!\times\!N$ identity matrix.
For $E\!<\!0$ the matrix $\MM^o$ is real and lookig for solutions of $\det(\MM)\!=\!0$ is equivalent to solve
\begin{equation}\label{ConditionFiniteNegative}
m^o_i(E)=-\alpha\qquad i=1,2,\cdots,N
\end{equation}
for each of the $N$ eigenvalues $m^o_i$ of $\MM^o$. Solutions of Eq.~\eqref{ConditionFiniteNegative} give real and negative energies of the MW bound states in the gas of scatterers.
For $E\!>\!0$ the situation is slightly different. A continuum of states is allowed for the MW, nevertheless, for a large enough number of scatterers, precursors of the bulk Bloch states of the infinite periodic system can be identified in the form of complex poles of the analytical continuation of $G$ to the lower-half plane of complex energies. In our approach, this corresponds to the fact that $\MM^o$ is now a complex matrix and the poles of the extended $G$ can be found by solving
\begin{equation}\label{ConditionFinitePositive}
m^o_i(z)=-\alpha\qquad i=1,2,\cdots,N
\end{equation}
for complex energies of the form $z\!=\!E-i\hbar\Gamma/2$, where $E$ and $\Gamma\!>\!0$ represents, respectively, position and band-width (i.e. inverse lifetime) of the eigenstate. The latter, in an extended and ordered system, would be a quasi-Bloch state, i.e. a state showing the periodicity properties of a Bloch one within the gas of scatterers but with a finite lifetime inside of it.

\section{Bravais lattices}\label{SecBravais}

\begin{figure}[tb]
\includegraphics[width=0.48\textwidth]{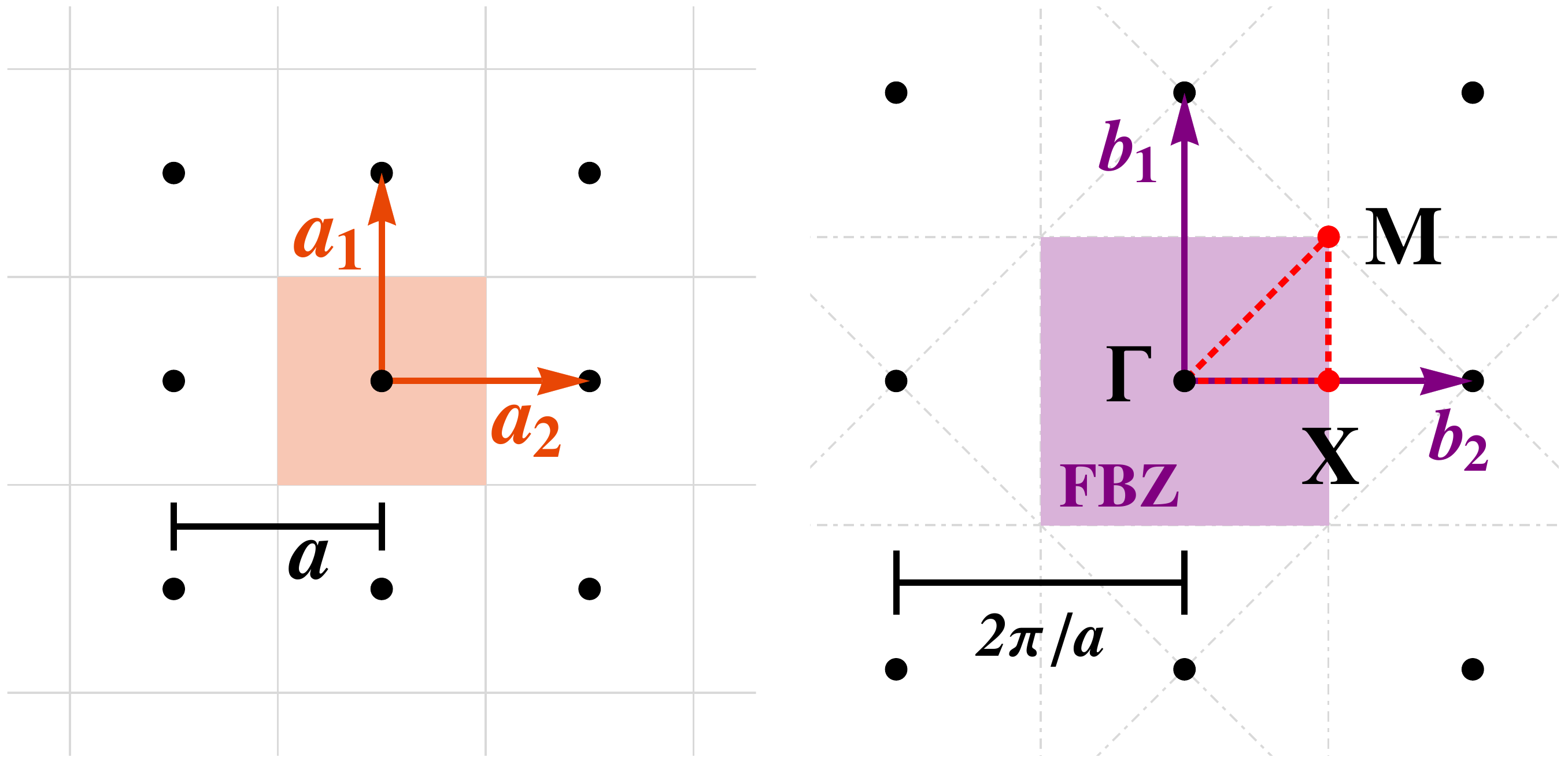}
\caption{(Color online)
Left: Schematic representation in real space of a square lattice of primitive vectors $\aaa_1\!=\!a(1,0)$ and $\aaa_2\!=\!a(0,1)$. The shadowed area represents the unit cell.
Right: Reciprocal lattice in $k$-space corresponding to the square lattice. Consequently the reciprocal primitive vectors are $\bbb_1\!=\!\frac{2\pi}{a}(1,0)$ and $\bbb_2\!=\!\frac{2\pi}{a}(0,1)$. The $1^{\rm st}$ Brillouin zone (FBZ) is shadowed. The high-symmetry points $\GGamma\!=\!(0,0)$, ${\bf X}\!=\!\frac{\pi}{a}(1,0)$, and ${\bf M}\!=\!\frac{\pi}{a}(1,1)$ are highlighted and the $\GGamma\!-\!{\bf X}\!-\!{\bf M}\!-\!\GGamma$ path (red, dashed) constitutes an irreducible symmetry path.}
\label{FigSchemeSquare}\end{figure}

\begin{figure}[tb]
\includegraphics[width=0.48\textwidth]{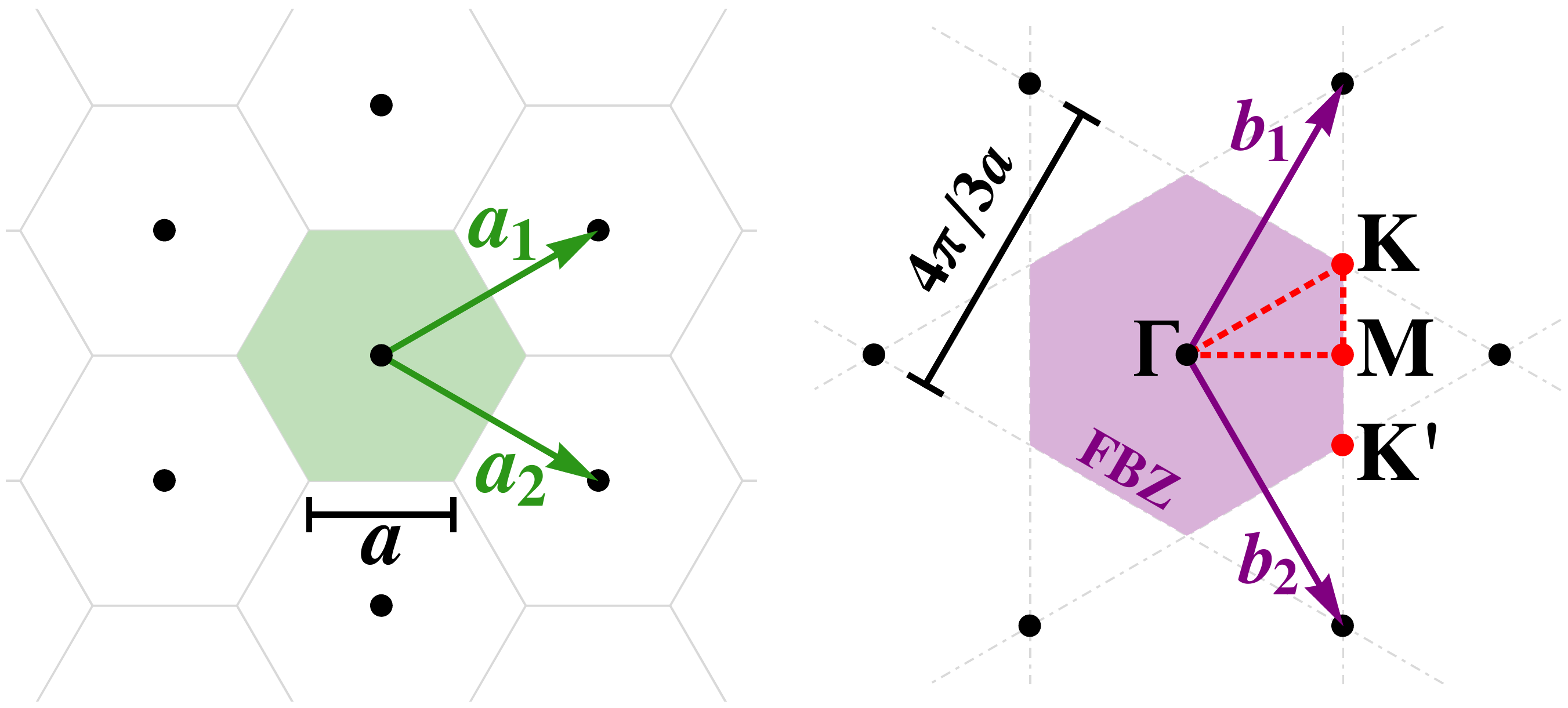}
\caption{(Color online)
Left: Real-space representation of a triangular lattice of primitive vectors $\aaa_1\!=\!\frac{3a}{2}(1,1/\sqrt{3})$ and $\aaa_2\!=\!\frac{3a}{2}(1,-1/\sqrt{3})$. The shadowed region represents the Wigner-Seitz unit cell.
Right: Reciprocal lattice for the triangular one, being  $\bbb_1\!=\!\frac{2\pi}{3a}(1,\sqrt{3})$ and $\bbb_2\!=\!\frac{2\pi}{3a}(1,-\sqrt{3})$ the reciprocal primitive vectors. The FBZ is shadowed and the high-symmetry points $\GGamma\!=\!(0,0)$, ${\bf M}\!=\!\frac{2\pi}{3a}(1,0)$, ${\bf K}\!=\!\frac{2\pi}{3a}(1,1/\sqrt{3})$, and ${\bf K'}\!=\!\frac{2\pi}{3a}(1,-1/\sqrt{3})$ are highlighted.
The $\GGamma\!-\!{\bf M}\!-\!{\bf K}\!-\!\GGamma$ path (red, dashed) is an high-symmetry one.}
\label{FigSchemeTriangular}
\end{figure}

In this section we adapt the general formalism introduced in Sec.~\ref{SecGeneral} to the case in which $B$ atoms are arranged in a Bravais lattice: an infinite periodic structure where a unit cell, containing only \emph{one} atom, is repeated to cover the entire 2D space (c.f. Figs.~\ref{FigSchemeSquare} and~\ref{FigSchemeTriangular}). Such a lattice is invariant under any translation $\RR\!\in\!L$ with $L\!=\!\{n_1{\bf a}_1+n_2{\bf a}_2:n_1,n_2\in {\mathbb Z}\}$, where the set $L$ is defined in terms of the two primitive vectors ${\bf a}_1$ and ${\bf a}_2$.
Consequently we can define the reciprocal lattice as a periodic structure invariant under translations $\KK\!\in\!RL$ with $RL=\{n_1{\bf b}_1+n_2{\bf b}_2:n_1,n_2\in {\mathbb Z}\}$, where $\bbb_1$ and $\bbb_2$ are the reciprocal primitive vectors, defined by the relation $\aaa_i\cdot\bbb_j\!=\!2\pi\delta_{ij}$ ($i,j\!=\!1,2$)\cite{KittelBook}.
Some examples of Bravais and corresponding reciprocal lattices are presented in Figs.~\ref{FigSchemeSquare} and~\ref{FigSchemeTriangular}.

The condition for the existence of an eigenstate, i.e. $\det(\MM)\!=\!0$, implies that the homogeneous system
\begin{equation}\label{HomogeneousSystem}
\sum_{i=1}^\infty \MM_{ji}D_i = 0 \qquad j=1,2,\cdots,\infty
\end{equation}
associated to the inhomogeneous one of Eq.~\eqref{LinearSystem} admits a non-trivial solution. Notice that now the number of scatterers $N$, and thus the number of equations and unknowns in the system, is infinite \cite{CarusottoPRA08AntezzaPRL09,AntezzaPRA09}.
In such a periodic structure Bloch's theorem holds, implying that
\begin{equation}\label{BlochTheoremBravais}
D_i=D_j e^{i\qq\cdot(\rr_j-\rr_i)},
\end{equation}
where $\qq$ is a vector of the $1^{st}$ Brillouin zone (FBZ) in reciprocal space. Resorting to this property all the equations of the homogeneous system~\eqref{HomogeneousSystem} become identical, so that the unique condition to verify is
\begin{equation}\label{ConditionBravaisReal}
\ln\left(\frac{e^\gamma}{2}\,k\atwo\right)-i\frac{\pi}{2}
+\sum_{\RR\in L^*} \frac{\pi\hbar^2}{m} g_0(\RR) e^{i\qq\cdot\RR}=0,
\end{equation}
where $L^*\!=\!L\setminus\{0\}$.
Due to the slow convergence of the sum in Eq.~\eqref{ConditionBravaisReal}, a rewriting of the equation in terms of reciprocal-lattice vectors is appropriate. The delicate details of this transformation are reported in Appendix~\ref{AppChangeOfSpaceDiag}, the result being
\begin{align}\label{ConditionBravaisReciprocal}
&C_\infty +\ln\left(\frac{e^\gamma}{2}\right) + \frac{2\pi}{\AAA}\frac{1}{k^2-q^2}
\nonumber\\
&\quad +\frac{2\pi}{\AAA} \sum_{\KK\in RL^*} \left(\frac{1}{k^2-|\KK-\qq|^2}+\frac{1}{K^2} \right)+\alpha=0.
\end{align}
Here $C_\infty$ is a coefficient depending only on the geometry of the Bravais lattice, its origin and definition follow from the real-to-reciprocal lattice transformation, presented in App.~\ref{AppChangeOfSpaceDiag}.
We also reintroduced the 2D interaction coefficient $\alpha\!=\!\ln(\atwo/a)$, ($a$ being an arbitrary unit of length) and $\AAA$ is the area of the real-space unit cell of the Bravais lattice.

\begin{figure}[tb]
\includegraphics[width=0.48\textwidth]{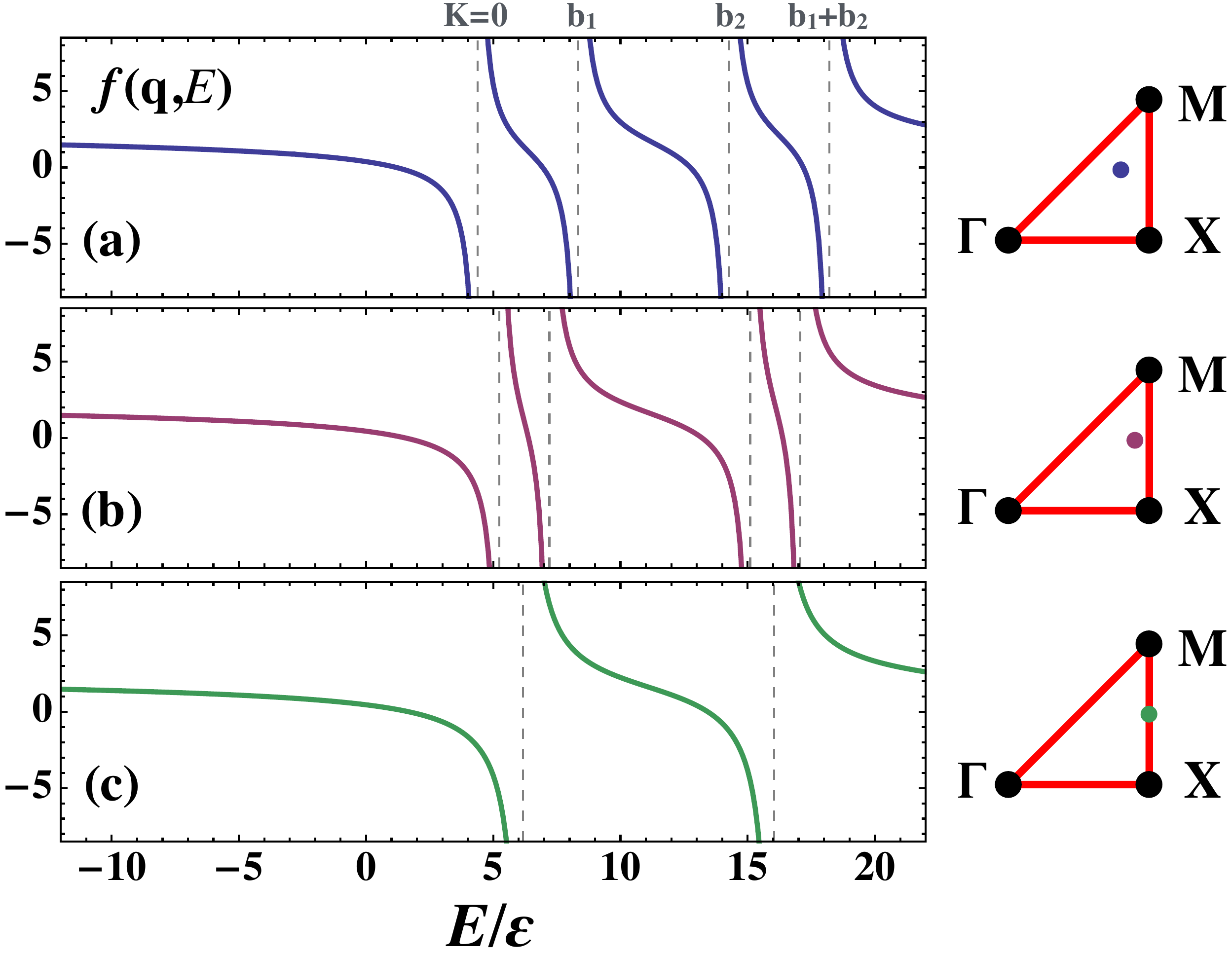}
\caption{(Color online) For a square lattice (cf. Fig.~\ref{FigSchemeSquare}), behavior of $f(\qq,E)$ as a function of $E$ for different values of $\qq$ (namely $\frac{\pi}{a}(0.8,0.5)$, $\frac{\pi}{a}(0.9,0.5)$, and $\frac{\pi}{a}(1,0.5)$ from top to bottom). Dashed, gray vertical lines represent the lowest values of $E_{\rm free}$ for the selected $\qq$.
In the sketches on the right of each panel a point indicates the corresponding $\qq$ inside the $\GGamma\!-\!{\bf X}\!-\!{\bf M}\!-\!\GGamma$ symmetry path.
Here $\varepsilon\!=\!\hbar^2/ma^2$.}
\label{FigDivergenceSquare}\end{figure}

It is worth noticing that Eq.~\eqref{ConditionBravaisReciprocal} can be cast as
\begin{equation}\label{ConditionBravaisReciprocalShort}
f(\qq,E)=-\alpha,
\end{equation}
with the introduction of the interaction-independent function
\begin{align}\label{Definitionf}
&f(\qq,E)=C_\infty +\ln\left(\frac{e^\gamma}{2}\right) + \frac{2\pi}{\AAA}\frac{1}{k^2-q^2}
\nonumber\\
&\quad +\frac{2\pi}{\AAA} \sum_{\KK\in RL^*} \left(\frac{1}{k^2-|\KK-\qq|^2}+\frac{1}{K^2} \right).
\end{align}
It follows by its definition that $f(\qq,E)$ diverges for
\begin{equation}\label{Definitionkfree}
k=|\KK-\qq|\qquad \forall\,\KK\in RL,
\end{equation}
where wave vector and energy of the MW are related by $E\!=\!\hbar^2k^2/2m$.
Furthermore $f(\qq,E)$ results monotonically decreasing in $E$ between two divergences, ensuring that for a given $\qq$ only \emph{one} solution of Eq.~\eqref{ConditionBravaisReciprocalShort} exists between them.
Another remarkable consequences is that if Eq.~\eqref{Definitionkfree} holds than condition~\eqref{ConditionBravaisReciprocalShort} is satisfied only for $|\alpha|\!\to\!\infty$, that is in the limit of non interaction between MW and scatterers (cf. Fig.~\ref{FigDivergenceSquare} for a practical example).
This implies that for $|\alpha|\!\to\!\infty$ one would recover the dispersion relation of a free MW, whose energy is given by
\begin{equation}\label{DefinitionEfree}
E_{\rm free}=\frac{\hbar^2}{2m}|\KK-\qq|^2\qquad \forall\,\KK\in RL.
\end{equation}

\subsection{Square lattice}\label{SecSquare}
As a first example of AAC we consider $B$ scaterers arranged in a square lattice of spacing $a$, for which the primitive vectors are simply $\aaa_1\!=\!a(1,0)$ and $\aaa_2\!=\!a(0,1)$. The reciprocal lattice and the FBZ are consequently defined, as illustrated in Fig.~\ref{FigSchemeSquare}.

\subsubsection{Infinite system}\label{SecSquareInfinite}
To study the band structure of the artificial atomic square lattice (AASL) one needs to solve Eq.~\eqref{ConditionBravaisReciprocal}, where in this case $C_\infty\!\simeq\!1.42646$ (cf. Eq.~\eqref{DefCinfty}).
It results convenient from a computational point of view to evaluate once for all the function $f(\qq,E)$ defined in Eq.~\eqref{Definitionf} and look afterwards for solutions of Eq.~\eqref{ConditionBravaisReciprocalShort} for a given $\alpha$.
In Fig.~\ref{FigDivergenceSquare} we plot $f(\qq,E)$ for different values of $\qq$ inside the FBZ. As expected $f$ diverges each time that $E\!=\!E_{\rm free}$ (i.e. when condition~\eqref{Definitionkfree} is satisfied), but when $\qq$ gets closer to the boundary of the FBZ some values of $E_{\rm free}$ corresponding to different $\KK$ eventually tend to each other. A solution of Eq.~\eqref{ConditionBravaisReciprocalShort} remains thus ``trapped'' in the small corridor formed by the two divergences and it tends to $E\!=\!E_{\rm free}$ in the degeneracy limit, independently of $\alpha$.

\begin{figure}[tb]
\includegraphics[width=0.48\textwidth]{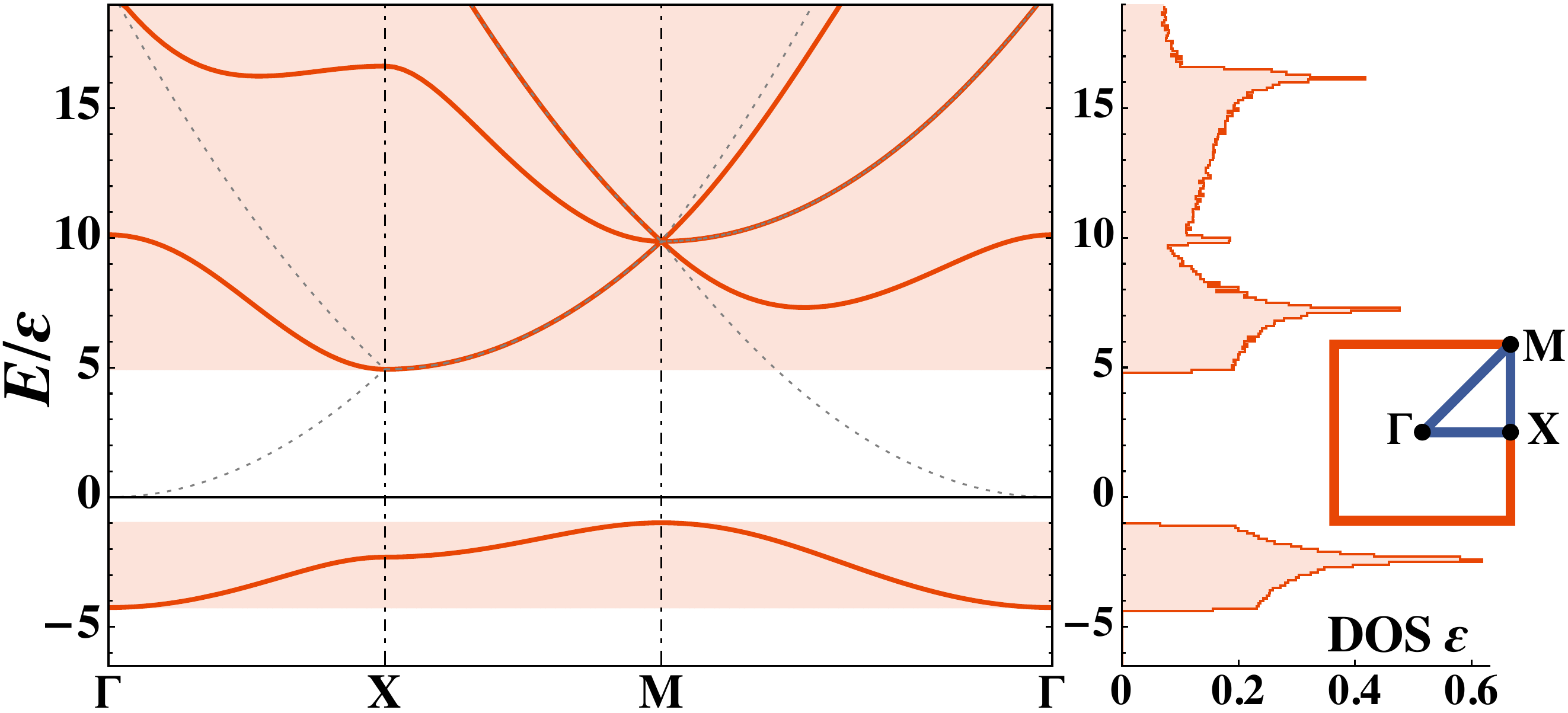}
\caption{(Color online) For the artificial atomic square lattice,
band structure (left) and density of states (right) evaluated at $\alpha\!=\!-0.75$.
Left: The four lowest energy bands are evaluated along the irreducible symmetry path $\GGamma\!-\!{\bf X}\!-\!{\bf M}\!-\!\GGamma$ within the $1^{\rm st}$ Brilouin zone (see inset).
Dashed, gray lines correspond to the energy spectrum of a free matter wave.
Right: Density of states (DOS) obtained by evaluating the energies of the same bands in $N_{\!s}\!\simeq\!6100$ points sampled inside the path. For each bin of the histogram (of width $\delta E\!=\!0.1\varepsilon$, $\varepsilon\!=\!\hbar^2/ma^2$) the band-normalized DOS is evaluated according to Eq.~\eqref{DefDOSNumerical}.}
\label{FigSpecterSquare}\end{figure}

We present in Fig.~\ref{FigSpecterSquare} the spectrum and density of states (DOS) of the AASL for $\alpha\!=\!-0.75$.
The spectrum is evaluated along the $\GGamma\!-\!{\bf X}\!-\!{\bf M}\!-\!\GGamma$ symmetry path (cf. inset).
The free-MW energy $E_{\rm free}$ is also shown and, as expected, only one solution of Eq.~\eqref{ConditionBravaisReciprocalShort} exists between two (eventually coinciding) free bands.
An omnidirectional gap is found in the band structure and, correspondingly, in the DOS.
The numerical evaluation of the DOS can be obtained by sampling each energy band in $N_s$ points within the FBZ. Subsequently the formula
\begin{equation}\label{DefDOSNumerical}
{\rm DOS}(E)=\frac{N_E}{N_s\delta E}
\end{equation}
gives the DOS in the energy range $(E,E+\delta E)$, being $N_E$ the number of sampled energies falling in the energy interval. The DOS defined in Eq.~\eqref{DefDOSNumerical} is normalized to the number of bands taken into account.
Finally we remark that the lowest and isolated band lays entirely at negative energies for this value of $\alpha$, meaning that the MW Bloch states are actually bound.

\begin{figure}[tb]
\includegraphics[width=0.48\textwidth]{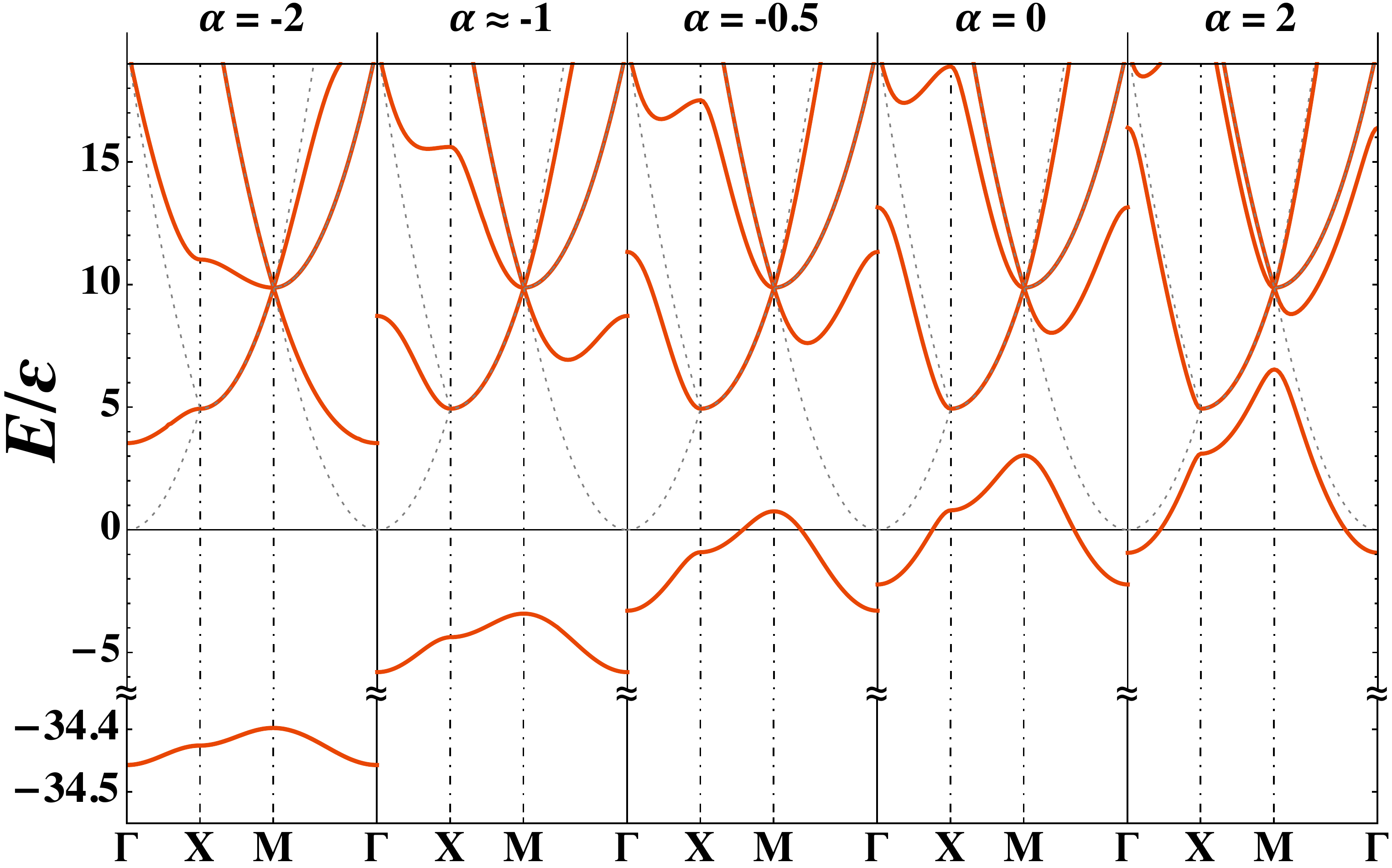}
\caption{(Color online)
Square lattice: band structure for different values of the interaction parameter $\alpha$. The spectra are evaluated along the same path shown in Fig.~\ref{FigSpecterSquare}.
Dashed, gray lines correspond to the energy spectrum of a free matter wave. Energies are normalized on $\varepsilon\!=\!\hbar^2/ma^2$. }
\label{FigSpecterComparedSquare}\end{figure}

A fundamental aspect of our model relies in the possibility to tune its physical properties by acting on $\alpha$. This remarkable feature shows up in Fig.~\ref{FigSpecterComparedSquare}, where the spectra of the AASL are compared for different values of $\alpha$.
The most evident modification concerns the lowest band, which results isolated for $\alpha\!\lesssim\!0.7453$, so that a gap opens in the spectrum. Furthermore the band becomes more and more deep and flat, leading to small group velocities for the corresponding eigenstates.
For $|\alpha|\!\gg\!1$ matter wave and scatterers are weakly interacting and, as expected, the band structure tends to the free-MW spectrum (grey dashed curves in Fig.~\ref{FigSpecterComparedSquare}).

\subsubsection{Finite-size effects}\label{SecSquareFinite}

\begin{figure}[tb]
\includegraphics[width=.48\textwidth]{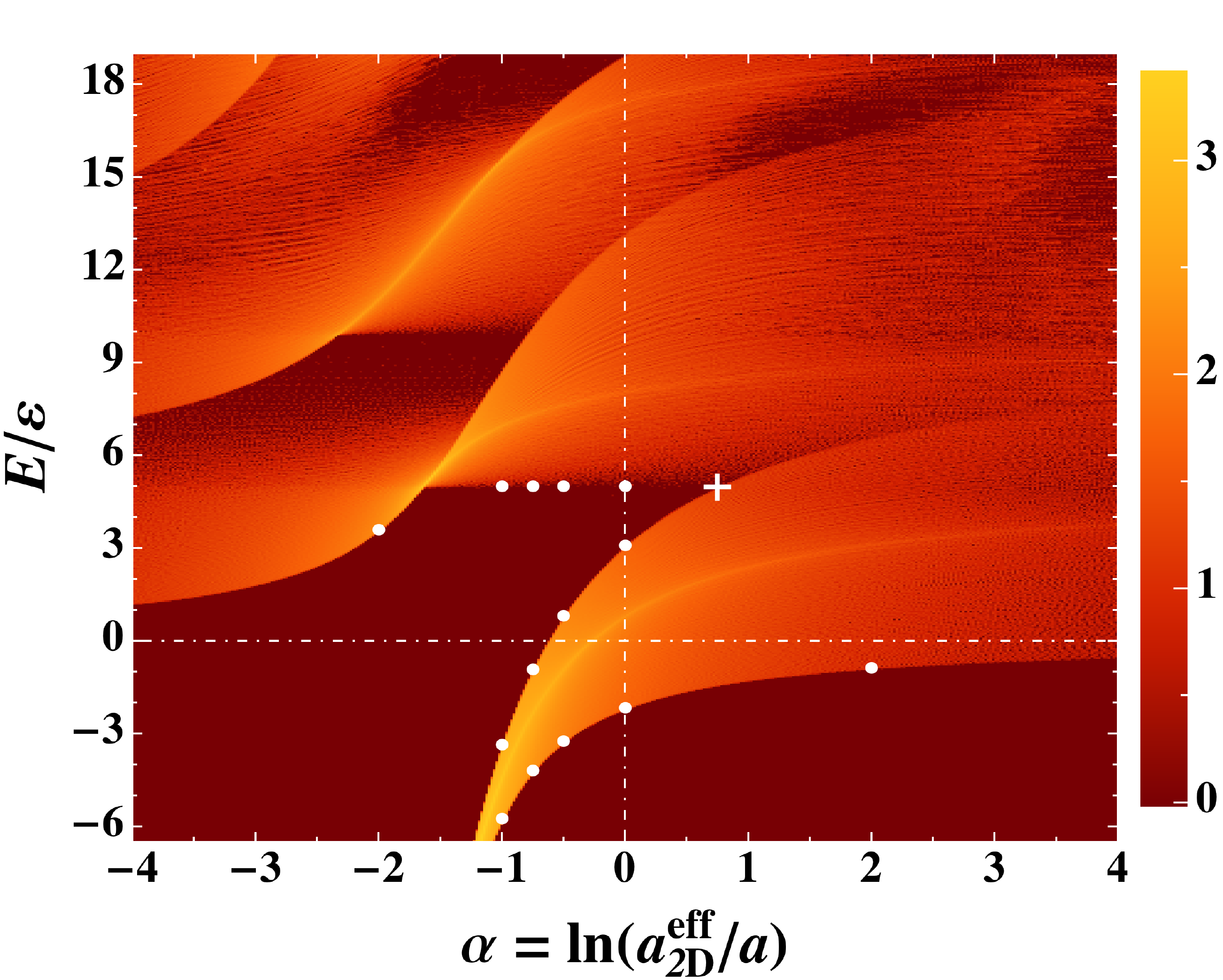}
\caption{(Color online)
Finite-sized square lattice: DOS per scatterer in the plane
$[\alpha,E/\varepsilon]$ for a system of $N\!=\!2551$ scatterers
arranged in a square lattice inside a disk of radius $R\!=\!26a$.
The energies are discretized with a step of $0.01\varepsilon$ ($\varepsilon\!=\!\hbar^2/ma^2$).
For given $E\!<\!0$ all the $N$ solutions of Eqs.~\eqref{ConditionFiniteNegative} are selected.
For $E\!>\!0$ the sampled values are used as starting points to find
$N$ complex poles of $G$ according to Eq.~\eqref{zFirstOrder}.
Only quasi-Bloch bulk states are selected as explained in the text.
The color-map is applied to 
$\log_{10}(\frac{N_{\rm p}}{N}\frac{\varepsilon}{\delta\alpha\,\delta E})$,
where $N_{\rm p}$ is the number of selected poles of $G$ within a
rectangular bin of area $\delta\alpha\,\delta E$
($\delta\alpha\!=\!0.02$ and $\delta E\!=\!0.05\varepsilon$).
$\bullet$ indicate the positions of band boundaries as expected from the
analysis of an infinite system (Figs.~\ref{FigSpecterSquare} and \ref{FigSpecterComparedSquare}).
Analogously the symbol $+$ marks the expected contact-point between the lowest and the first excited band, corresponding to the gap closure.}
\label{FigFiniteSizeSquare}\end{figure}

Up to now we considered ideal periodic systems, but for both theoretical and practical interests it is crucial to investigate the robustness of the above-reported results for realistic finite-size systems.
In this case one can not resort to Bloch's theorem and the use of the general approach presented in Sec.~\ref{SecGeneral} is necessary.
In a typical experimental setup, atomic clouds can be manipulated in optical lattices extending over $\sim\!60$ sites per dimension.
This means that 2D artificial atomic lattices with $\sim\!10^3$ trapped $B$-scatterers are experimentally realizable.
In Fig.~\ref{FigFiniteSizeSquare} we thus present the density of states on the $[\alpha,E]$ plane for a finite system of $\sim\!2500$ scatterers arranged in a square lattice inside a circular region of radius $R\!=\!26a$.
The finite-size results are here compared with those of an ideal infinite system and the agreement is excellent.

\begin{figure}[tb]
\includegraphics[width=0.48\textwidth]{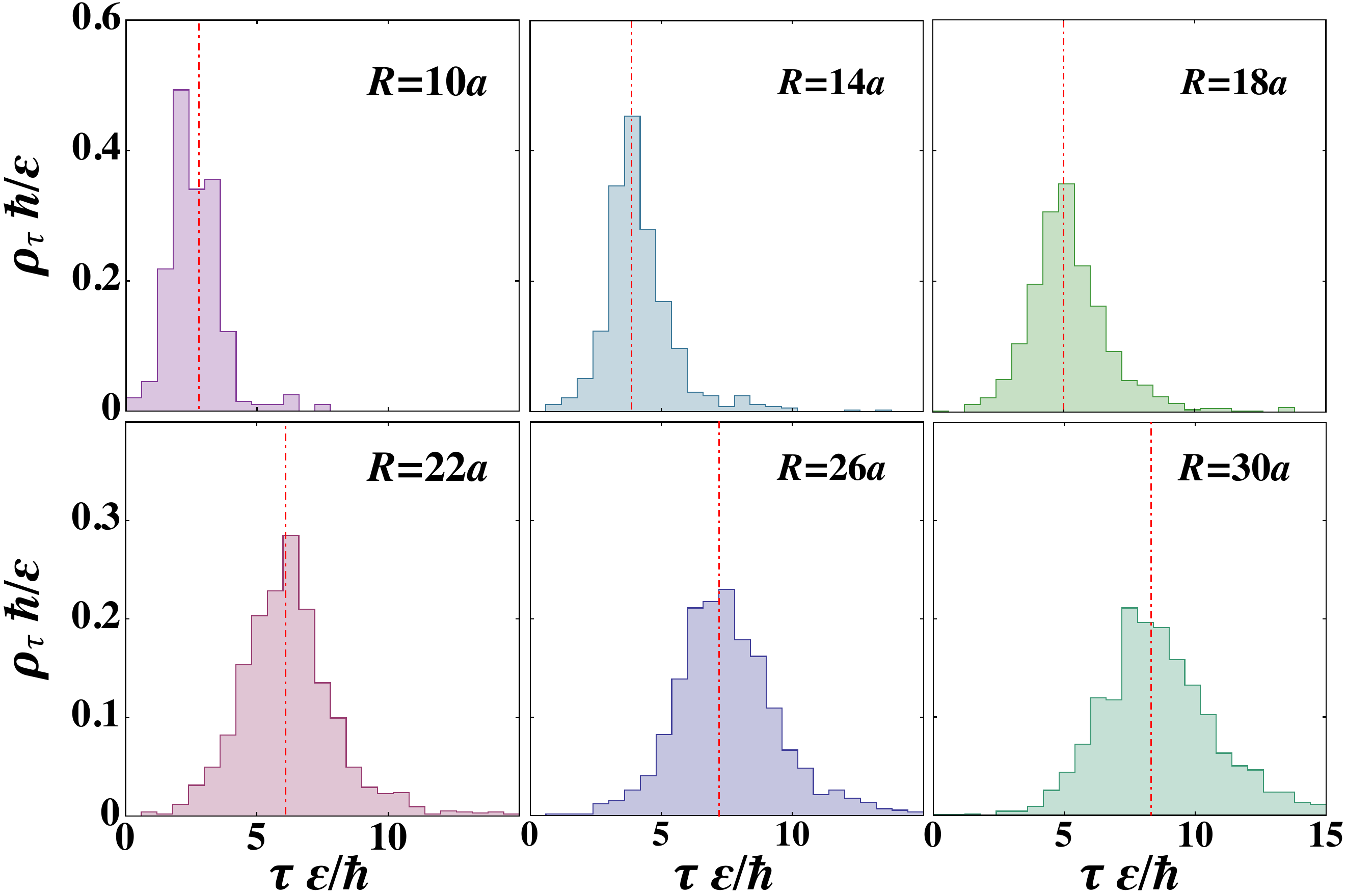}
\caption{(Color online)
Square lattice of finite size: normalized densities of life-times $\tau$ for $E/\varepsilon\!=\!6.5\pm0.5$ and $\alpha\!=\!-0.75\pm0.1$ ($\varepsilon\!=\!\hbar^2/ma^2$).
Different panels correspond to different radii $R$ of the scattering region. Dot-dashed red vertical lines indicate in each panel $R/\bar{v}_g$, where $\bar{v}_g$ is the average group velocity in the selected region of the $[\alpha,E]$ plane.}
\label{FigTauDistributionSquare}
\end{figure}

In the evaluation of the DOS of Fig.~\ref{FigFiniteSizeSquare} it results computationally convenient to fix $E$ and correspondingly look for values of $\alpha$ satisfying conditions~\eqref{ConditionFiniteNegative} or~\eqref{ConditionFinitePositive}.
For $E\!<\!0$ the $N\!\times\!N$ real symmetric matrix $\MM^o(E)$ and its eigenvalues can be evaluated, thus the $\alpha$ solving Eqs.~\eqref{ConditionFiniteNegative} are immediately obtained.
For $E\!>\!0$ one can use $\tilde{E}\!=\!E$ as starting point to find an approximate solution of the complex Eqs.~\eqref{ConditionFinitePositive} \cite{AntezzaPRA10}. We write $z\!=\!\tilde{E}+\delta z$ and we assume the $1^{\rm st}$ order expansion $m_i^o(z)\!=\!m_i^o(\tilde{E})+\delta z\,{m_i^o}'(\tilde{E})$ to be valid.
By choosing $\alpha\!=\!-\Re[m_i^o(\tilde{E})]$ each equation reduces to $i\Im[m_i^o(\tilde{E})]+\delta z\,{m_i^o}'(\tilde{E})\!=\!0$, from which $\delta z$ can be directly derived.
Energies $E$ and band-width $\Gamma$ of the quasi-Bloch states follow from
\begin{equation}\label{zFirstOrder}
z=E-i\frac{\hbar}{2}\Gamma\simeq\tilde{E}-i\frac{\Im[m_i^o(\tilde{E})]}{{m_i^o}'(\tilde{E})}.
\end{equation}
Notice that the derivative ${m_i^o}'(\tilde{E})$ can be evaluated resorting to the generalized Hellmann-Feynman theorem: for the complex symmetric matrix $\MM^o$ one has $dm_i^o(z)/dz\!=\!\vec{u}_i\cdot [d\MM^o(z)/dz]\,\vec{u}_i$, where $\vec{u}_i$ is the right eigenvector of eigenvalue $m_i^o(z)$, normalized as $\vec{u}_i\cdot\vec{u}_i\!=\!1$ (instead of the usual $\vec{u}_i\cdot\vec{u}_i^*\!=\!1$).

\begin{figure}[tb]
\includegraphics[width=0.42\textwidth]{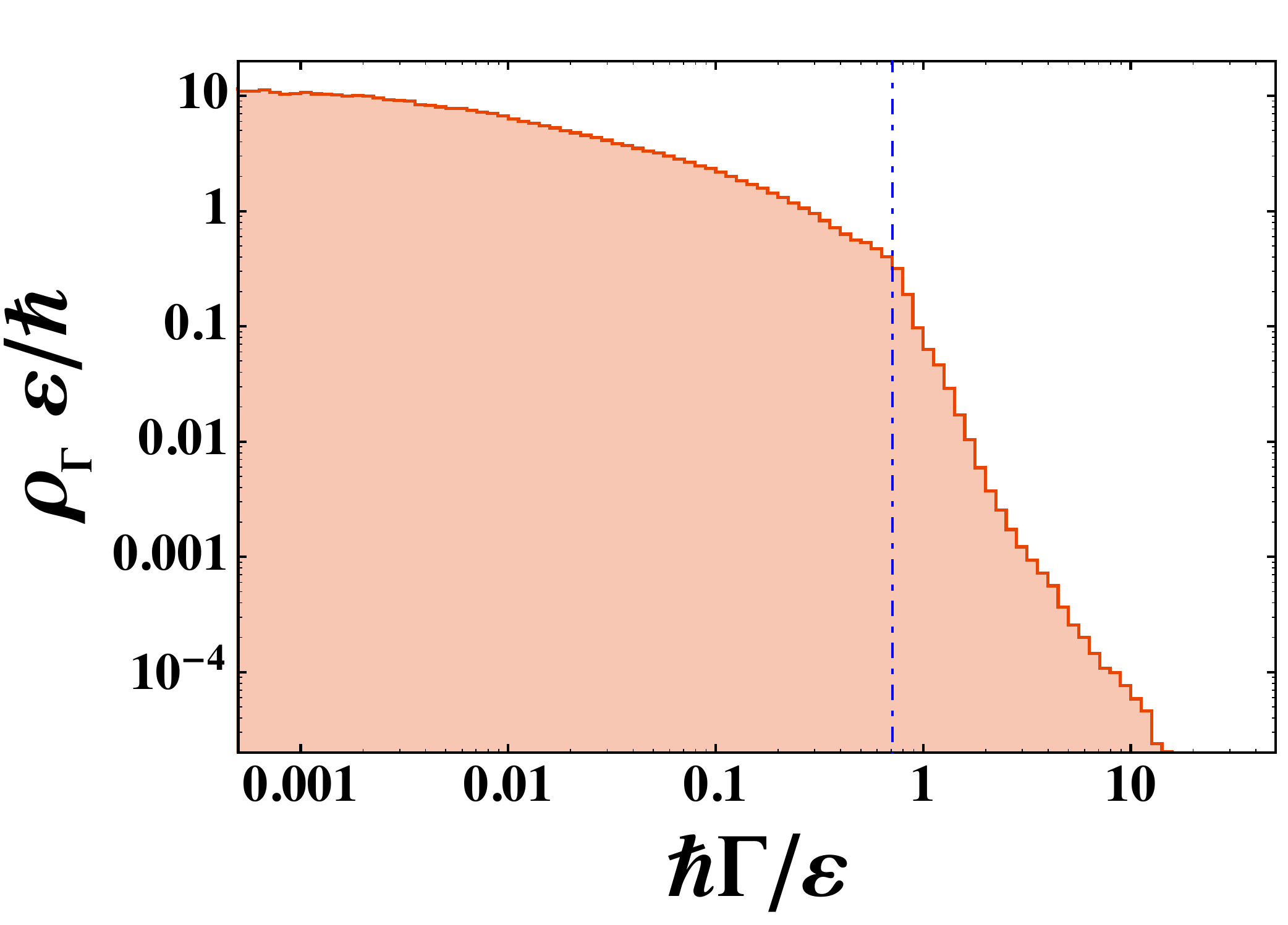}
\caption{(Color online) 
For the same finite-size square lattice considered in Fig.~\ref{FigFiniteSizeSquare}: normalized density of bandwidths $\Gamma$ for all the positive-energy poles of the Green function ($\varepsilon\!=\!\hbar^2/ma^2$).
The dot-dashed vertical blue line at $\Gamma\!=\!\Gamma_{\rm max}$ marks the change in behavior of the density $\rho_\Gamma$. Solutions with $\Gamma\!>\!\Gamma_{\rm max}$ have been rejected because incompatible with the behavior of a quasi-Bloch bulk state.
Here we found $\Gamma_{\rm max}\!\simeq\!0.7\varepsilon/\hbar$ (i.e. $\Gamma_{\rm max}\!\simeq\!2$kHz for a matter wave of $^{87}$Rb atoms in an AASL with $a\!=\!500{\rm nm}$).}
\label{FigGammaSquare}
\end{figure}

The $1^{\rm st}$ order approximation~\eqref{zFirstOrder} results sufficient if $\Im[m_i^o(\tilde{E})]$ is small \cite{AntezzaPRA10} and this is the case for most of the solutions we find.
In particular for quasi-Bloch states one expects a lifetime $\tau\!=\!1/\Gamma\!\simeq\!R/v_g$, where $R$ is the radius of the region containing the scatterers and $v_g$ the group velocity of the state. To verify this behavior we selected a small window on the $[\alpha,E]$ plane, studying the distributions of $\tau$ when the system-size varies. The results of this analysis are presented in Fig.~\ref{FigTauDistributionSquare}. We find that the distribution peak follows the expected behavior, confirming that the first-order approximation is sufficient to individuate quasi-Bloch states.
Nevertheless, when the poles of $G$ are evaluated by Eq.~\eqref{zFirstOrder}, one can eventually find some results for which the $1^{\rm st}$ order is insufficient. These solutions would present non-physical negative values of $\Gamma$ and needs to be rejected (in the case presented in Fig.~\ref{FigFiniteSizeSquare} they constituted the 10.6\% of positive-energy states).
Furthermore, in a finite-size system, states others that quasi-Bloch ones can appear (such as edge-states) for which the law $\Gamma\!\simeq\!v_g/R$ is not valid. In order to tell them apart one can look at the density of $\Gamma$, shown in Fig.~\ref{FigGammaSquare} for the same system considered in Fig.~\ref{FigFiniteSizeSquare}.
A neat change in the behavior is found in $\Gamma_{\rm max}\!\sim\!v_{g\,\rm max}/R$, where $v_{g\,\rm max}$ is the highest estimated group velocity in the range of $E$ and $\alpha$ considered.
States with $\Gamma\!>\!\Gamma_{\rm max}$ can be in turn rejected, finally leaving only solutions behaving as quasi-Bloch bulk states (in the case of Fig.~\ref{FigFiniteSizeSquare} this led to the exclusion of an additional 8.4\% of solutions).
A natural question may arise concerning the dependence of $v_{g\,\rm max}$ on $\alpha$ and $E$. It has been verified that by considering a cut-off depending on $\alpha$ and $E$ results are qualitatively the same as those obtained using the aforementioned selection method.

\subsubsection{Introduction and effects of disorder}\label{SecSquareDisorder}

\begin{figure}[tb]
\includegraphics[width=0.48\textwidth]{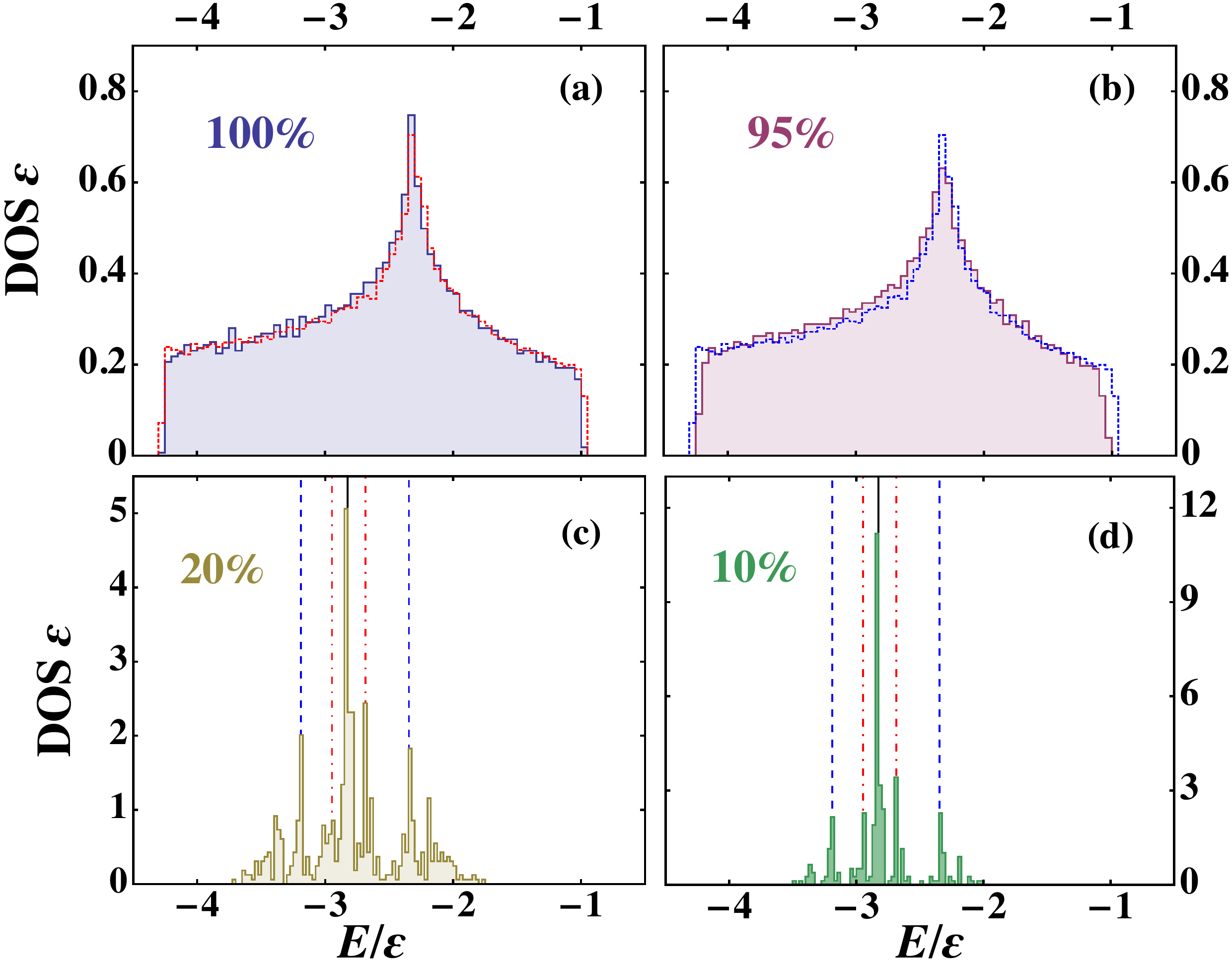}
\caption{(Color online)
Disordered square lattice: normalized negative-energy DOS at $\alpha\!=\!-0.75$ for an AASL of radius $R\!=\!32a$, corresponding to 3209 available lattice sites. Different panels refers to different percentage of randomly occupied sites: 100\% in~(a), 95\% in~(b), 20\% in~(c), 10\% in~(d).
Histograms are obtained with a bin size $\delta E\!=\!0.05\varepsilon$ in~(a)-(b) and $\delta E\!=\!0.025\varepsilon$ in~(c)-(d).
Superimposed dashed histograms in panels~(a)-(b) represent the DOS for an infinite periodic system.
The vertical lines in panels~(c)-(d) indicate the energies of few-body bound states: $AB$ dimer (solid black), $AB_2$ trimer with $B$ atoms separated by $a$ (dashed blue) and $a\sqrt{2}$ (dot-dashed red).
Here $\varepsilon\!=\!\hbar^2/ma^2$.}
\label{FigDisorderSquare}
\end{figure}

A remarkable advantage of our atomic artificial lattices with respect to other one-species models is the possibility to naturally introduce disorder in the system. Loading the $B$-atom optical lattice with non-unitary filling would result in the presence of randomly unoccupied sites constituting random defects in the artificial crystal \cite{AntezzaPRA13}.
Our general theoretical approach, presented in Sec.~\ref{SecGeneral}, consents to investigate the effects of this kind of disorder.
In particular we show in Fig.~\ref{FigDisorderSquare} the negative-energy DOS of an AASL of radius $R\!=\!32a$ at different filling factors, obtained fixing $\alpha\!=\!-0.75$.
In panel~(a) 100\% of sites are occupied and finite-size effects can be investigated by comparing the DOS with that of the periodic system (already presented in Fig.~\ref{FigSpecterSquare}). There are no significant discrepancies between the two quantities, confirming the robustness of the results with respect to the system size.
The DOS in presence of 5\% of vacant sites, panel~(b), appears in turns compatible with the results for an ideal periodic system, thus proving robustness also against a small amount of vacancies.
For a large number of unoccupied sites, instead, the periodicity of the lattice gets lost and the matter wave interacts locally with few-body clusters of scatterers. As can be seen in panels~(c) and~(d) of Fig.~\ref{FigDisorderSquare}, this gives rise to a DOS which is more and more peaked around the energies of few-body $AB_n$ bound states.
The energies of these bound states can be derived from Eqs.~\eqref{ConditionFiniteNegative}, where the eigenvalues of $\MM^o$ can be analytically obtained for $n\!\le\!4$. The explicit expressions can be found in Eqs.~(45) and~(46) of \cite{AntezzaPRA10}, where the emergence of disorder-localized states for low filling have been investigated.

\subsection{Triangular lattice}\label{SecTriangular}
Another relevant example of atomic artificial Bravais lattice is the triangular one, which analysis is also preliminary to the study of intriguing non-Bravais structures, such as graphene and the kagom\'e lattice.
In this case the primitive vectors are $\aaa_1\!=\!a(\sqrt{3},1)\sqrt{3}/2$ and $\aaa_2\!=\!a(\sqrt{3},-1)\sqrt{3}/2$,
from which the real and reciprocal lattice, illustrated in Fig.~\ref{FigSchemeTriangular}, are defined.

\subsubsection{Infinite system}
To investigate the spectral properties of the atomic artificial triangular lattice (AATL) we need, once again, to solve Eq.~\eqref{ConditionBravaisReciprocalShort}, now for $C_\infty\!\simeq\!0.959662$.
We can thus proceed as done in Sec.~\ref{SecSquareInfinite} for the square lattice.

\begin{figure}[tb]
\includegraphics[width=0.48\textwidth]{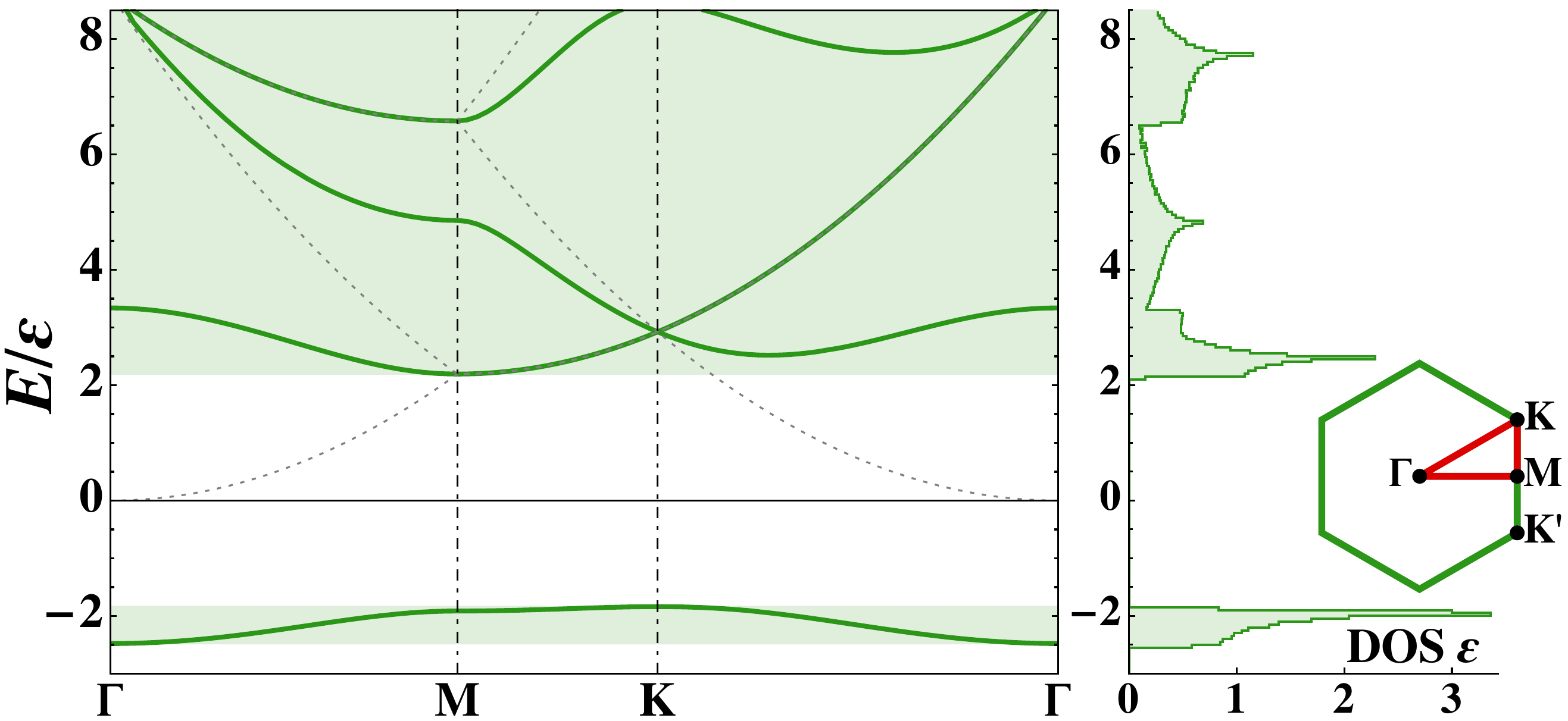}
\caption{(Color online)
Artificial atomic triangular lattice: band structure (left) and DOS (right) for $\alpha\!=\!-0.6$.
Left: Behavior of the four lowest energy bands along the $\GGamma\!-\!{\bf M}\!-\!{\bf K}\!-\!\GGamma$ path (see inset to the right). Dashed gray lines indicates the dispersion relation for a free matter wave.
Right: DOS, as defined in Eq.~\eqref{DefDOSNumerical}, obtained by evaluating the energy of the bands in $N_s\!=\!7600$ points sampled inside the aforementioned symmetry path. The bin-size of the histogram is $\delta E\!=\!0.05\varepsilon$ ($\varepsilon\!=\!\hbar^2/ma^2$).}
\label{FigSpecterTriangular}
\end{figure}

\begin{figure}[tb]
\includegraphics[width=0.48\textwidth]{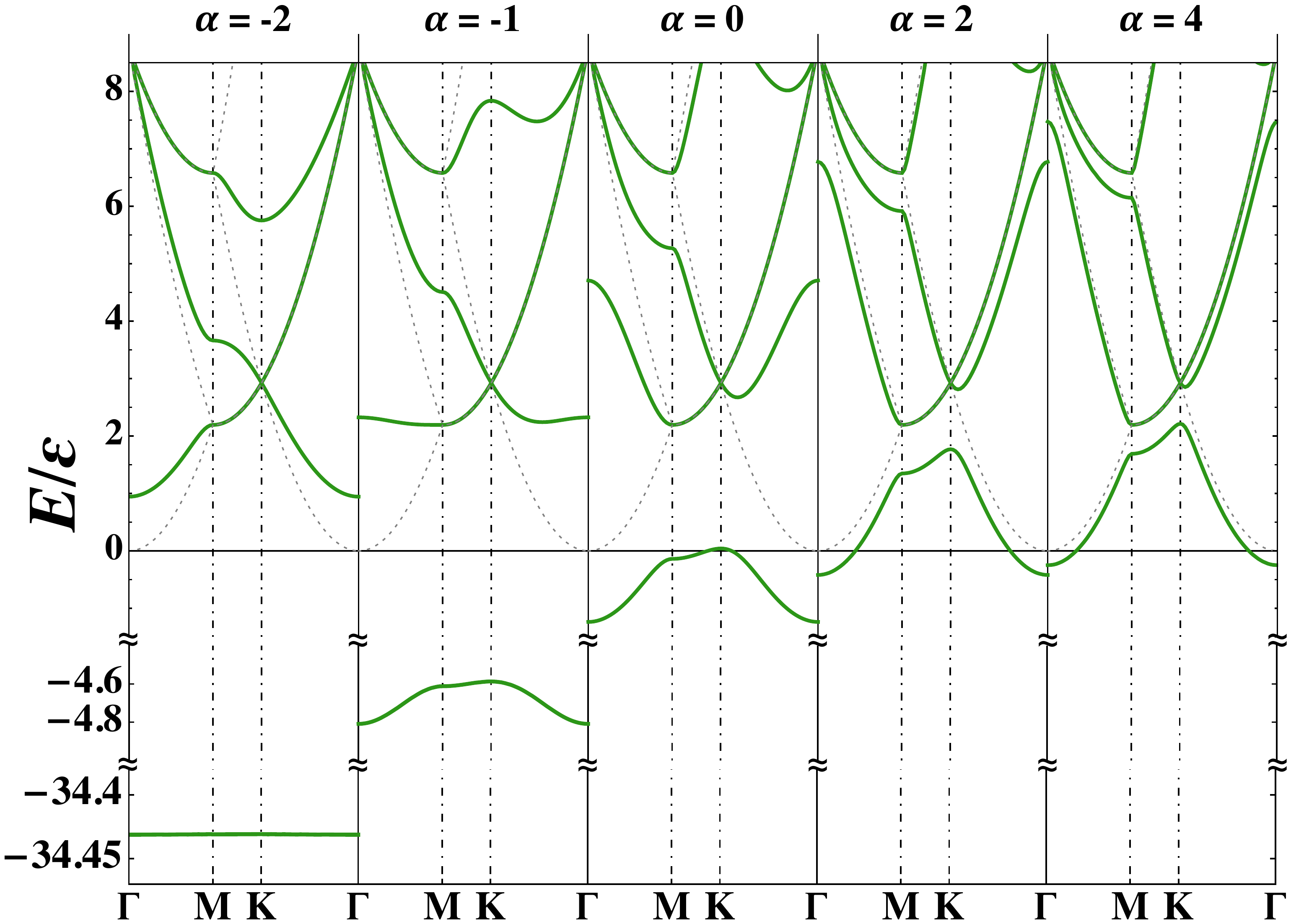}
\caption{(Color online)
Triangular lattice: comparison of the band structure for different values of the interaction parameter $\alpha$.
The spectra-evaluation path is the same of Fig.~\ref{FigSpecterTriangular}.
Dashed, gray lines correspond to the energy spectrum of a free matter wave.
Energies are normalized on $\varepsilon\!=\!\hbar^2/ma^2$.}
\label{FigSpecterComparedTriangular}
\end{figure}

In Fig.~\ref{FigSpecterTriangular} we present a typical spectrum of the system, evaluated in $\alpha\!=\!-0.6$.
The band structure, studied along the $\GGamma\!-\!{\bf M}\!-\!{\bf K}\!-\!\GGamma$ high-symmetry path (see inset), shows the presence of a gap, which the DOS confirms to be omnidirectional.
Also in this case we verify the existence of a single solution of Eq.~\eqref{ConditionBravaisReciprocalShort} between two values of $E_{\rm free}$.
The versatility of this artificial lattice emerges in Fig.~\ref{FigSpecterComparedTriangular}, where we compare its spectra for different values of the interaction parameter $\alpha$.
One finds that there exists a gap of tunable width for $\alpha\!\lesssim\!3.853$ and, again, the lowest and isolated band gets fast thin and deep in energy by decreasing $\alpha$.

\subsubsection{Finite-size and disorder effects}

\begin{figure}[tb]
\includegraphics[width=0.48\textwidth]{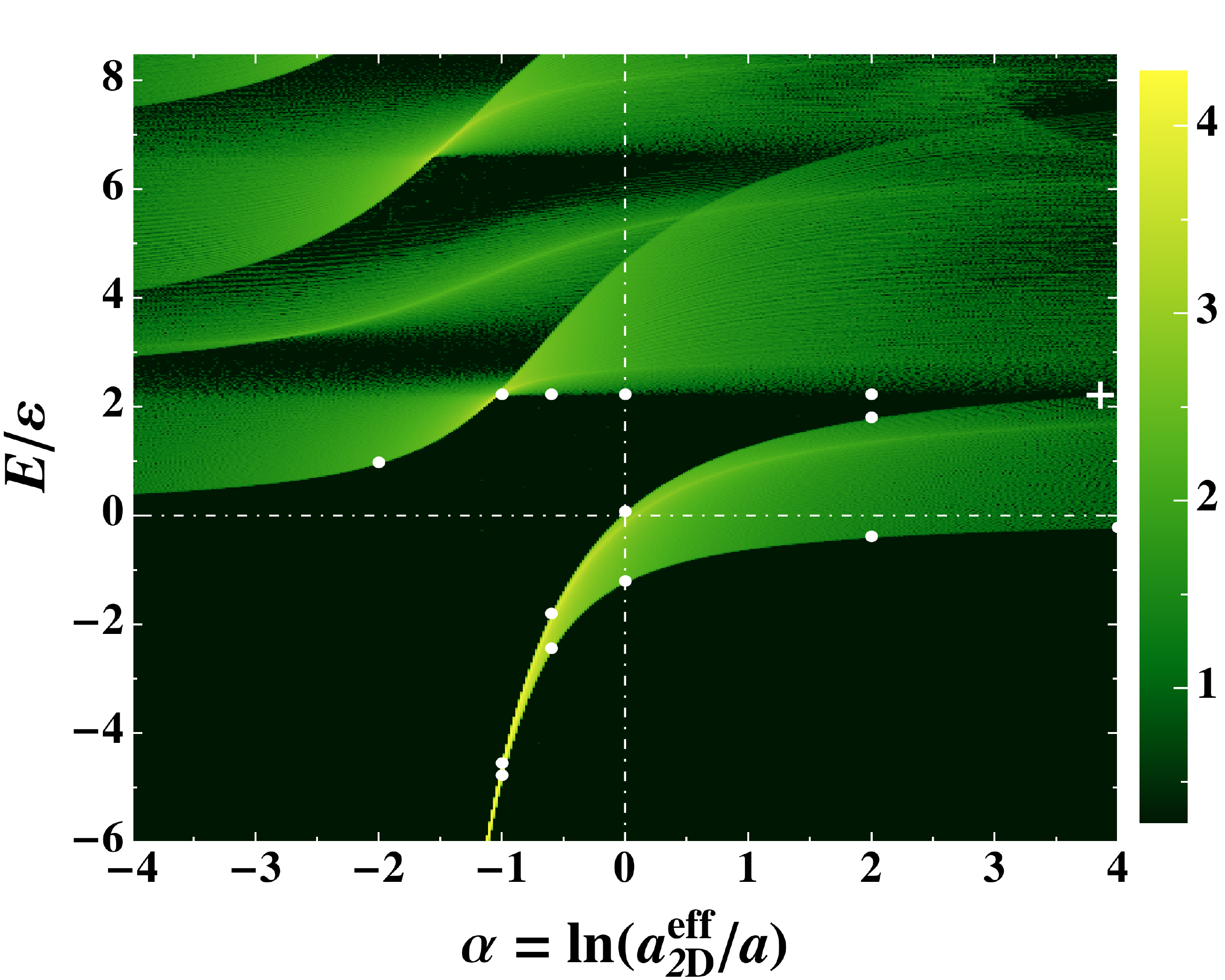}
\caption{(Color online)
Triangular lattice of finite size: DOS per scatterer in the plane
$[\alpha,E/\varepsilon]$ for a system of $N\!=\!2125$ scatterers
arranged in a triangular lattice inside a disk of radius $R\!=\!42a$.
The same method of Fig.~\ref{FigFiniteSizeSquare} has been used, but discretizing $E$ in steps of $0.005\varepsilon$ ($\varepsilon\!=\!\hbar^2/ma^2$).
Positive-energy quasi-Bloch bulk states have been selected imposing $\Gamma\!<\!\Gamma_{\rm max}\!\simeq\!0.3\varepsilon/\hbar$ (i.e. $\Gamma_{\rm max}\!\simeq\!880$Hz for a matter wave of $^{87}$Rb atoms in an AATL with $a\!=\!500{\rm nm}$).
The color-map is applied to the quantity
$\log_{10}(\frac{N_{\rm p}}{N}\frac{\varepsilon}{\delta\alpha\,\delta E})$,
where $N_{\rm p}$ is the number of selected poles of $G$ within a
rectangular bin of area $\delta\alpha\,\delta E$
($\delta\alpha\!=\!0.02$ and $\delta E\!=\!0.025\varepsilon$).
$\bullet$ indicate the expected positions of the band boundaries as evaluated for the infinite system (Figs.~\ref{FigSpecterTriangular} and \ref{FigSpecterComparedTriangular}).
Analogously the $+$ marks the expected contact-point between the lowest and the first excited band.}
\label{FigFiniteSizeTriangular}
\end{figure}

For sake of completeness we test again our results against finite-size effects.
With the same method illustrated in Sec.~\ref{SecSquareFinite} we evaluated the DOS on the plane $[\alpha,E]$ for an experimental-size system of $\sim\!2100$ scatterers arranged in a triangular lattice inside a circular region of radius $R\!=\!42a$. The results, shown in Fig.~\ref{FigFiniteSizeTriangular}, are once again in good agreement with predictions based on the analysis of an ideal AATL.
We also verified that, in analogy to the square-lattice case, results are robust if a small amount of vacancies is randomly introduced in the triangular structure, while few body states becomes dominant for low fillings.

\section{Non-Bravais lattices}\label{SecNonBravais}

\begin{figure}[tb]
\includegraphics[width=0.48\textwidth]{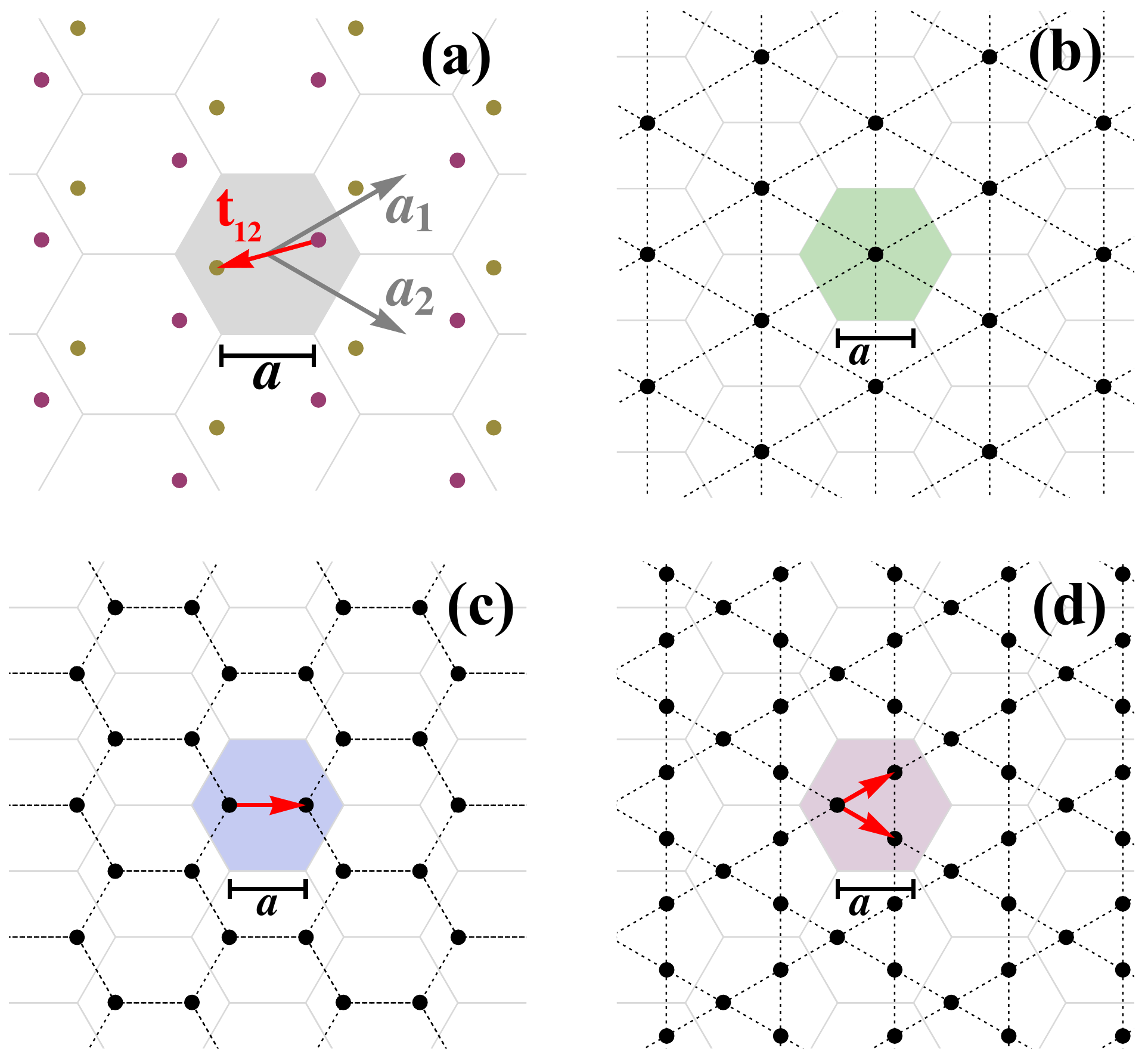}
\caption{(Color online)
Panel (a): Schematic representation of a $M\!=\!2$ non-Bravais lattice.
It can be seen as a triangular lattice (Fig.~\ref{FigSchemeTriangular}) with 2 atoms per unit cell (shaded hexagon) whose relative position is $\ttt_{12}$. Equivalently it can be obtained as the superposition of two interpenetrating triangular lattices (distinguished by colors) displaced by $\ttt_{12}$ with respect to each other.
The generating vectors $\aaa_{1,2}\!=\!\frac{3a}{2}(1,\pm1/\sqrt{3})$ are also indicated.
Panel (b): Representation of the triangular lattice as example of basic Bravais lattice, i.e. with \emph{one} atom per unit cell. The distance between nearest atoms is $a\sqrt{3}$.
Panel (c): Representation of the hexagonal lattice of graphene, a \emph{two}-atom non-Bravais lattice based on the triangular one, obtained for $\ttt_{12}\!=\!(\aaa_1+\aaa_2)/3\!=\!(a,0)$. With these definitions, the side of the hexagons has length $a$.
Panel (d): Representation of the kagom\'e lattice, a \emph{three}-atom non-Bravais lattice based on the triangular one, obtained for $\ttt_{12}\!=\!\aaa_1/2$ and $\ttt_{13}\!=\!\aaa_2/2$. The nearest-neighbor distance results $a\sqrt{3}/2$.}
\label{FigSchemeNB}
\end{figure}

This section is devoted to generalize the formalism introduced in Sec.~\ref{SecBravais} to the case in which $B$ scatterers are arranged in a non-Bravais lattice: an infinite periodic structure in which a unit cell, containing now $M$ atoms, is repeated to cover the 2D plane. Such a structure can be equivalently seen as a set of $M$ identical Bravais lattices, the $m^{\rm th}$ and $n^{\rm th}$ being displaced by $\ttt_{mn}$ with respect to each other. The structure will be still invariant if translated by $\RR\!\in\!L$, for $L\!=\!\{n_1{\bf a}_1+n_2{\bf a}_2:n_1,n_2\in {\mathbb Z}\}$, where ${\bf a}_{1,2}$ are the primitive vectors of a sub-lattice (cf. panel (a) of Fig.~\ref{FigSchemeNB}). All the properties of a Bravais lattice stay separately valid for each sub-lattice.
If we denote by $\RR_i$ the central position of the $i^{\rm th}$ unit cell and by $\rrho_m$ the position of the $m^{\rm th}$ atom with respect to this reference, the location of a scatterer in the non-Bravais lattice is given by $\rr_{im}\!=\!\RR_i+\rrho_m$. The linear system~\eqref{HomogeneousSystem} becomes
\begin{equation}\label{HomogeneousSystemNB1}
\sum_{i=1}^\infty \sum_{m=1}^M \MM_{jn,im}\,D_{im} = 0
\end{equation}
for each $j=1,2,\cdots,\infty$ and $n=1,2,\cdots,M$. Practically each index has been split with respect to previous notation (i.e. $j\to jn$ and $i\to im$) such that the first index runs on the lattice cells while the second indicates at which of the $M$ sub-lattices the scatterer belongs. Accordingly the definition of $\MM$ given in Eq.~\eqref{DefinitionM} still holds.
Bloch's theorem is now separately valid for each sub-lattice, so that
\begin{equation}\label{BlochTheoremNonBravais}
D_{im}=D_{jm}\, e^{i\qq\cdot(\rr_{jm}-\rr_{im})}=D_{jm}\, e^{i\qq\cdot(\RR_j-\RR_i)},
\end{equation}
while $D_{im}$ and $D_{jn}$ stay independent for $m\!\neq\!n$ \cite{AntezzaPRA09}. This assumption makes the present treatment not valid in the few accidental cases in which the non-Bravais lattice degenerates in a Bravais one.
For an arbitrary choice of $j$, all the equations of system~\eqref{HomogeneousSystemNB1} take the form
\begin{equation}\label{HomogeneousSystemNB2}
\sum_{m=1}^M D_{jm} \sum_{i=1}^\infty\,\MM_{jn,im}\, e^{i\qq\cdot(\RR_j-\RR_i)}=0,
\end{equation}
giving an homogeneous system of $M$ equations in the $M$ unknowns $D_{jm}$, whose matrix $\TT$ is defined by
\begin{equation}\label{DefinitionT}
\TT_{nm}=\sum_{i=1}^\infty\,\MM_{jn,im}\, e^{i\qq\cdot(\RR_j-\RR_i)}.
\end{equation}
The matrix elements read explicitly
\begin{align}
\label{DefinitionTOffDiagReal}
&\TT_{nm}=\sum_{R\in L} \frac{\pi\hbar^2}{m}\, g_0(\RR+\ttt_{nm})\,
e^{i\qq\cdot\RR} \qquad\text{for  } n\!\neq\!m
\\
\label{DefinitionTDiagReal}
&\TT_{nn}=\ln\left(\frac{e^\gamma}{2}\,k\atwo\right)-i\frac{\pi}{2}
+\sum_{\RR\in L^*} \frac{\pi\hbar^2}{m}\, g_0(\RR)\, e^{i\qq\cdot\RR},
\end{align}
with $\ttt_{nm}\!=\!\rrho_n\!-\!\rrho_m$. Condition $\det(\TT)\!=\!0$ has to be satisfied so that system~\eqref{HomogeneousSystemNB2} has a solution, which in turns means that this is the condition for the existence of an eigenstate of the $A$-atom MW in presence of a non-Bravais lattice of $B$ scatterers. 
Notice that the diagonal terms are all the same and that they correspond to the LHS of Eq.~\eqref{ConditionBravaisReal}. This naturally implies that for $M\!=\!1$ we get back to the case of a Bravais lattice: $\det(\TT)\!=\!\TT_{11}\!=\!0$ is exactly Eq.~\eqref{ConditionBravaisReal}.
It is again convenient to re-write the sums in terms of reciprocal-lattice vectors. Following the procedure described in App.~\ref{AppChangeOfSpace} we finally have
\begin{align}
\label{DefinitionTOffDiagReciprocal}
\TT_{nm}=&\frac{2\pi}{\AAA} \sum_{\KK\in RL} \frac{e^{i(\KK-\qq)\cdot\ttt_{nm}}}{k^2-|\KK-\qq|^2},
\\
\label{DefinitionTDiagReciprocal}
\TT_{nn}=&\,C_\infty +\ln\left(\frac{e^\gamma}{2}\right) + \frac{2\pi}{\AAA}\frac{1}{k^2-q^2}
\nonumber\\
&+\frac{2\pi}{\AAA} \sum_{\KK\in RL^*} \left(\frac{1}{k^2-|\KK-\qq|^2}+\frac{1}{K^2} \right)+\alpha.
\end{align}
Besides, from the latter expressions the matrix $\TT$ results explicitly hermitian.

\subsection{Graphene}\label{SecGraphene}
Considered the increasing interest towards the intriguing properties of graphene and its artificial realizations \cite{GeimNature07,CastroNetoRMP09,PoliniNature13}, it turns out natural to apply our model to investigate the behavior of the atomic artificial graphene (AAG)  (cf. \cite{BartoloEPL14} for more details):
a cold-atom quantum simulator of graphene where, differently from preexisting models presented in the field \cite{ZhuPRL07,TarruellNature12}, the periodic potential felt by the matter wave is generated by other atoms.
The honeycomb structure is actually a non-Bravais lattice resulting from the superposition of two interpenetrating triangular ones displaced by $\ttt\!=\!(a,0)$, as illustrated in Fig.~\ref{FigSchemeNB} panel (c).

\begin{figure}[tb]
\includegraphics[width=0.48\textwidth]{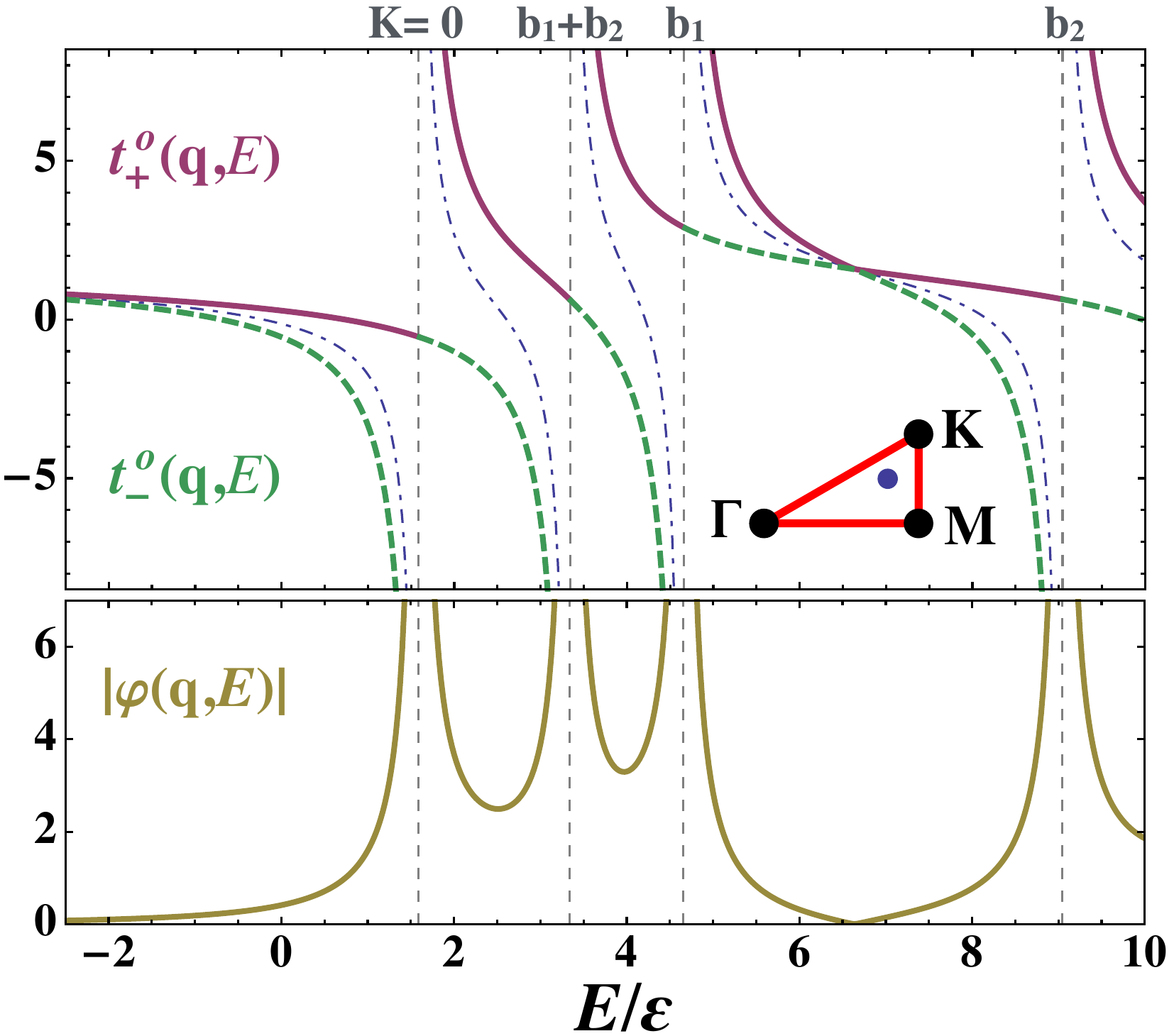}
\caption{(Color online) For the atomic artificial graphene.
Top: Behavior of $t^o_\pm(\qq,E)$ (solid purple for $+$ and green dashed for $-$) as a function of $E$ for $\qq\!=\!\frac{2\pi}{3a}(0.8,0.3)$. Dashed, gray vertical lines indicates the lowest values of $E_{\rm free}$ for the selected $\qq$.
Dot-dashed blue curves show the behavior of $f(\qq,E)$.
In the inset is shown the position of $\qq$ within the $\GGamma\!-\!{\bf M}\!-\!{\bf K}\!-\!\GGamma$ symmetry path (Fig.~\ref{FigSchemeTriangular}).
Bottom: Behavior of $|\varphi(\qq,E)|$ for the same $\qq$.
Here $\varepsilon\!=\!\hbar^2/ma^2$.}
\label{FigDivergenceGraphene}
\end{figure}

As usual we begin by considering an ideal infinite system.
For a generic two-atom non-Bravais lattice the condition for the existence of an eigenstate is $\det(\TT)\!=\!0$, being the matrix $\TT$ a $2\!\times\!2$ hermitian one. Once again the diagonal interaction term $\alpha$ can be isolated by writing $\TT\!=\!\TT^o+\II\alpha$, where $\II$ is the $2\!\times\!2$ identity matrix and
\begin{equation}\label{DefinitionTzero}
\TT^o=\TT(\alpha\!=\!0)=\begin{pmatrix}
f(\qq,E) &\varphi(\qq,E)\\
\varphi^*(\qq,E) &f(\qq,E)
\end{pmatrix},
\end{equation}
with $f(\qq,E)$ defined as in Eq.~\eqref{Definitionf} and
\begin{equation}\label{DefinitionPhi}
\varphi(\qq,E)=\frac{2\pi}{\AAA} \sum_{\KK\in RL} \frac{e^{i(\KK-\qq)\cdot\ttt}}{k^2-|\KK-\qq|^2}.
\end{equation}
It follows that condition $\det(\TT)\!=\!0$ is equivalent to
\begin{equation}\label{ConditionNonBravais}
t^o_\pm(\qq,E)=f(\qq,E)\pm|\varphi(\qq,E)|=-\alpha
\end{equation}
where $t^o_\pm$ are the two real eigenvalues of $\TT^o$.
It can be easily proved that when $f$ diverges, i.e. when $E\!\to\!E_{\rm free}$, the absolute value of $\varphi$ exactly cancels the divergence so that the left limit of $t^o_+$ is finite and equal to the right limit of $t^o_-$:
\begin{equation}
\lim_{E\to E_{\rm free}^\mp} t^o_\pm(\qq,E)=
\lim_{E\to E_{\rm free}} f(\qq,E) - \frac{\pi\hbar^2}{m\mathcal{A}}\,\frac{1}{E-E_{\rm free}}.
\end{equation}
A numerical example of that is given in Fig.~\ref{FigDivergenceGraphene}, where the two eigenvalues $t^o_\pm(\qq,E)$ and $|\varphi(\qq,E)|$ are plotted as functions of $E$ for a fixed $\qq$. In virtue of the aforementioned properties, only \emph{one} solution of Eq.~\eqref{ConditionNonBravais} exists between two solutions of the corresponding Eq.~\eqref{ConditionBravaisReciprocalShort}. This practically means that a band of a non-Bravais lattice with $M\!=\!2$ is always included between two bands of the corresponding Bravais one ($M\!=\!1$).

\begin{figure}[tb]
\includegraphics[width=0.48\textwidth]{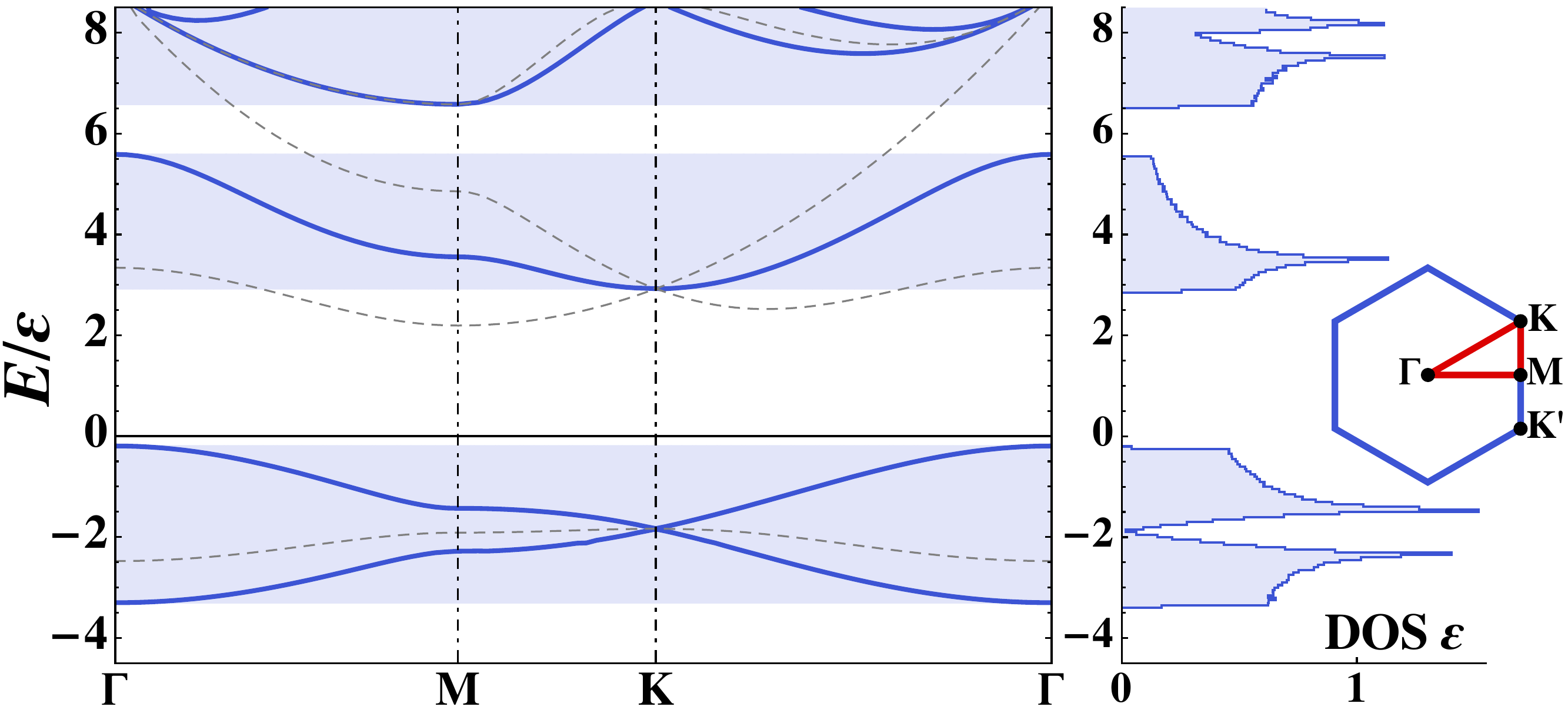}
\caption{(Color online)
Honeycomb lattice: band structure (left) and DOS (right) for the AAG at $\alpha\!=\!-0.6$.
Left: Behavior of the five lowest energy bands along the $\GGamma\!-\!{\bf M}\!-\!{\bf K}\!-\!\GGamma$ symmetry path (see inset on the right). Dashed gray curves show the band structure for the corresponding triangular lattice.
Right: DOS (defined in Eq.~\eqref{DefDOSNumerical}) obtained by sampling the bands energy in $N_s\!=\!7600$ points within the symmetry path. The histogram bin-size is $\delta E\!=\!0.05\varepsilon$ ($\varepsilon\!=\!\hbar^2/ma^2$).}
\label{FigSpecterGraphene}
\end{figure}

\begin{figure}[tb]
\includegraphics[width=0.48\textwidth]{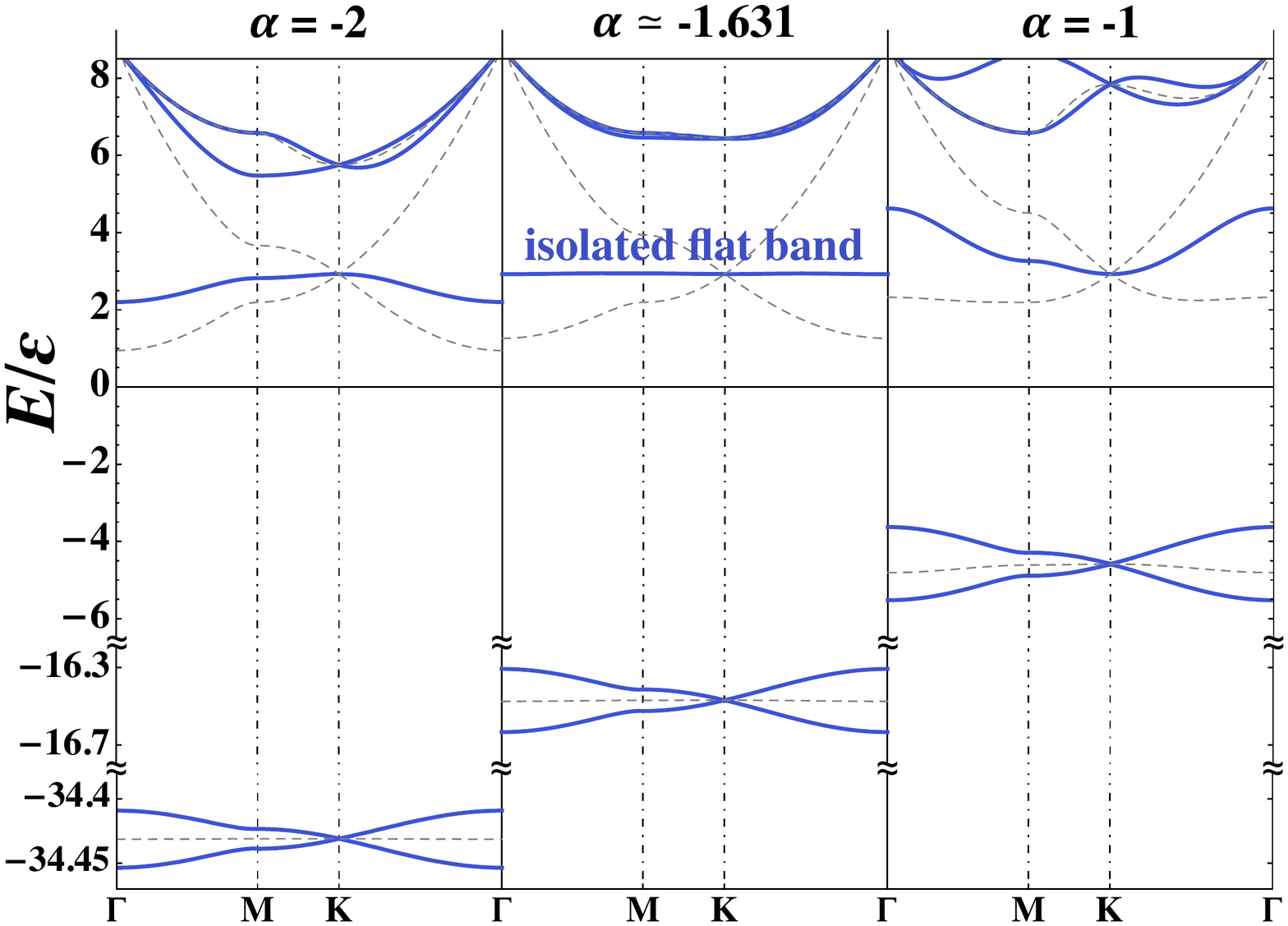}
\caption{(Color online)
Atomic artificial graphene: comparison of the band structures for different values of $\alpha$ evaluated along the $\GGamma\!-\!{\bf M}\!-\!{\bf K}\!-\!\GGamma$ symmetry path (see inset of Fig.~\ref{FigSpecterGraphene}).
For $\alpha\!\simeq\!-1.631$ the third and isolated band results topologically flat.
Dashed gray curves show the band structure for the corresponding triangular lattices.
The unit of energy is $\varepsilon\!=\!\hbar^2/ma^2$.}
\label{FigSpecterGrapheneFlat}
\end{figure}

An example of spectrum for the AAG is presented in Fig.~\ref{FigSpecterGraphene} for $\alpha\!=\!-0.6$.
Band structure and DOS point out the presence of a double gap, which can be manipulated and eventually closed acting on $\alpha$.
A key feature is the existence of two nonequivalent Dirac points, situated in $\KK,\KK'\!=\!\frac{2\pi}{3a}(1,\pm1/\sqrt{3})$ within the FBZ. Here the two lowest bands touch each other, creating a conic-shaped energy-momentum dispersion.
The matter-wave obeys so to a Dirac-like equation for relativistic massless fermions in which the role of the speed of light is played by $v_g$: the modulus of the group velocity of the wave along the cone.
Remarkably  the characteristics of the cones can be manipulated by acting on $\alpha$.
At first for $\alpha\!\lesssim\!0$ the cone lays at negative energies, so that relativistic physics is there played by bound states. Furthermore, similarly to the Bravais artificial lattices, the lowest bands gets flat by decreasing $\alpha$ leading to a widening of the cones (cf. Fig.~\ref{FigSpecterGrapheneFlat}). This results in a sensitive decrease of $v_g$, even below 1mm/s for typical experimental setups (i.e. up to $10^{-9}$ the corresponding quantity for electrons in real graphene).

Another striking feature emerges in $\alpha\!=\!\alpha_{\rm flat}\!\simeq\!-1.631$, values at which the third isolated band results topologically flat.
As shown in Fig.~\ref{FigSpecterGrapheneFlat}, for $\alpha\!>\!\alpha_{\rm flat}$ the third-band maximum is in $\GGamma$ and the minimum in ${\bf K}$, while these roles are exchanged for $\alpha\!<\!\alpha_{\rm flat}$.
Correspondingly the band concavity changes at the transition point, leaving the band completely flat around $E\!\simeq\!2.924\varepsilon$ ($\varepsilon\!=\!\hbar^2/ma^2$).
It may be argued that also in the case of Bravais artificial atomic lattices quasi-flat bands emerge for large and negative values of $\alpha$ (see lowest bands in Figs.~\ref{FigSpecterComparedSquare} and~\ref{FigSpecterComparedTriangular}), but some important differences exist between these bands and the one we find for the AAG. At first graphene's band is topologically flat, which means that $v_g\!=\!0$ for every value of $\qq$. The quasi-flat bands keep instead their structure, even though they are compressed in a small range of energies (this phenomenon appears clearly in Fig.~\ref{FigSpecterComparedSquare}). Moreover graphene's flat band lies at positive energies, in the propagative region of the matter wave.
The interest towards such kind of flat bands arises from their non-dispersivity: for non-interacting $A$ atoms (case considered in this work) any MW state would result stationary and localized, being the group velocity strictly zero on the band. The effects of an eventual $A\!-\!A$ interaction, even if extremely small, would be enhanced, leading to the emergence of strongly correlated phases \cite{WuPRL07PRB08}. Non-isolated flat bands have been recently observed in honeycomb lattices for polaritons \cite{JacqminPRL14}.

Finally the use of an optical lattice allows to manipulate and distort the hexagonal structure, thanks to the precise experimental control achievable on the potential landscape. In one-component artificial graphenes such distortion of the structure is expected leading to the motion and eventual merging of the Dirac cones \cite{CitMotionDirac}, a phenomenon recently observed experimentally for a Fermi gas in an honeycomb optical lattice \cite{TarruellNature12}.
Similarly such a conductor-to-insulator transition is expected to occur in our atomic artificial graphene.
In conclusion the analysis of finite-size and disorder effects shows the robustness of the aforementioned results against the system size.
Dirac cones and flat band persist for up to 2\% of vacant sites.

\subsection{Kagom\'e lattice}\label{SecKagome}

As an example of $M\!=\!3$ non-Bravais lattice we consider the kagom\'e one, motivated by an increasing interest toward this peculiar structure in which several phenomena of geometric frustration have been predicted \cite{CitFrustration}.
Like for graphene, the kagom\'e lattice is based on the triangular one, but now with \emph{three} atoms per unit cell forming an equilateral triangle, as depicted in Fig.~\ref{FigSchemeNB} panel (d).

\begin{figure}[tb]
\includegraphics[width=0.48\textwidth]{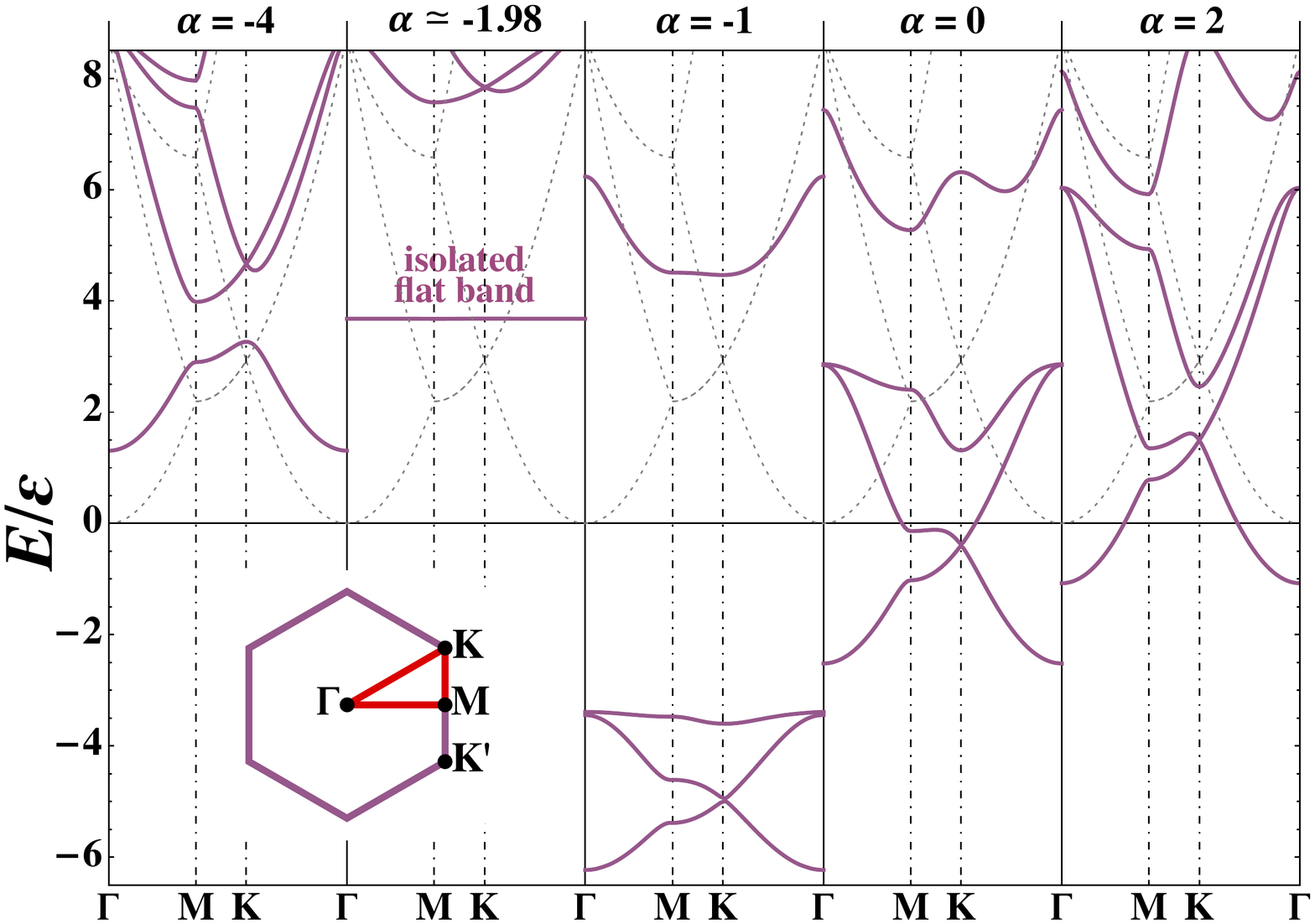}
\caption{(Color online)
Kagom\'e lattice: comparison of the band structures of the artificial lattice for different values of $\alpha$ evaluated along the $\GGamma\!-\!{\bf M}\!-\!{\bf K}\!-\!\GGamma$ symmetry path within the FBZ (see inset).
The two lowest bands touch in $\qq\!=\!\KK,\KK'$, forming a Dirac cone.
For $\alpha\!\simeq\!-1.98$ the fourth and isolated band results topologically flat.
For $\alpha\!<\!-1$ the three lowest band lay entirely at $E\!<\!6.5\varepsilon$ ($\varepsilon\!=\!\hbar^2/ma^2$)  and are not shown in the plot.
Dashed, gray lines correspond to the energy spectrum of a free matter wave.}
\label{FigSpecterComparedKagome}
\end{figure}

For the study of the periodic system, the condition $\det(\TT)\!=\!0$ translates now in looking for solutions of $t^o_i(\qq,E)\!=\!-\alpha$ with $i\!=\!1,2,3$ and $t^o_i$ being the three eigenvalues of the matrix $\TT^o$, defined in Eq.~\eqref{DefinitionTzero}.
Typical spectra for different values of $\alpha$ are presented in Fig.~\ref{FigSpecterComparedKagome}.
The two lowest bands present again two Dirac cones for $\qq\!=\!\KK,\KK'$ but, differently from the case of graphene, in the artificial kagom\'e lattice the third band moves together with the lowest two. The fourth band results flat and isolated for $\alpha\!\simeq\!-2$.
We notice again that for $|\alpha|\!\gg\!1$ the band structure approaches that of the free MW, as expected being the $A\!-\!B$ interaction weak in this limit.

\begin{figure}[tb]
\includegraphics[width=0.48\textwidth]{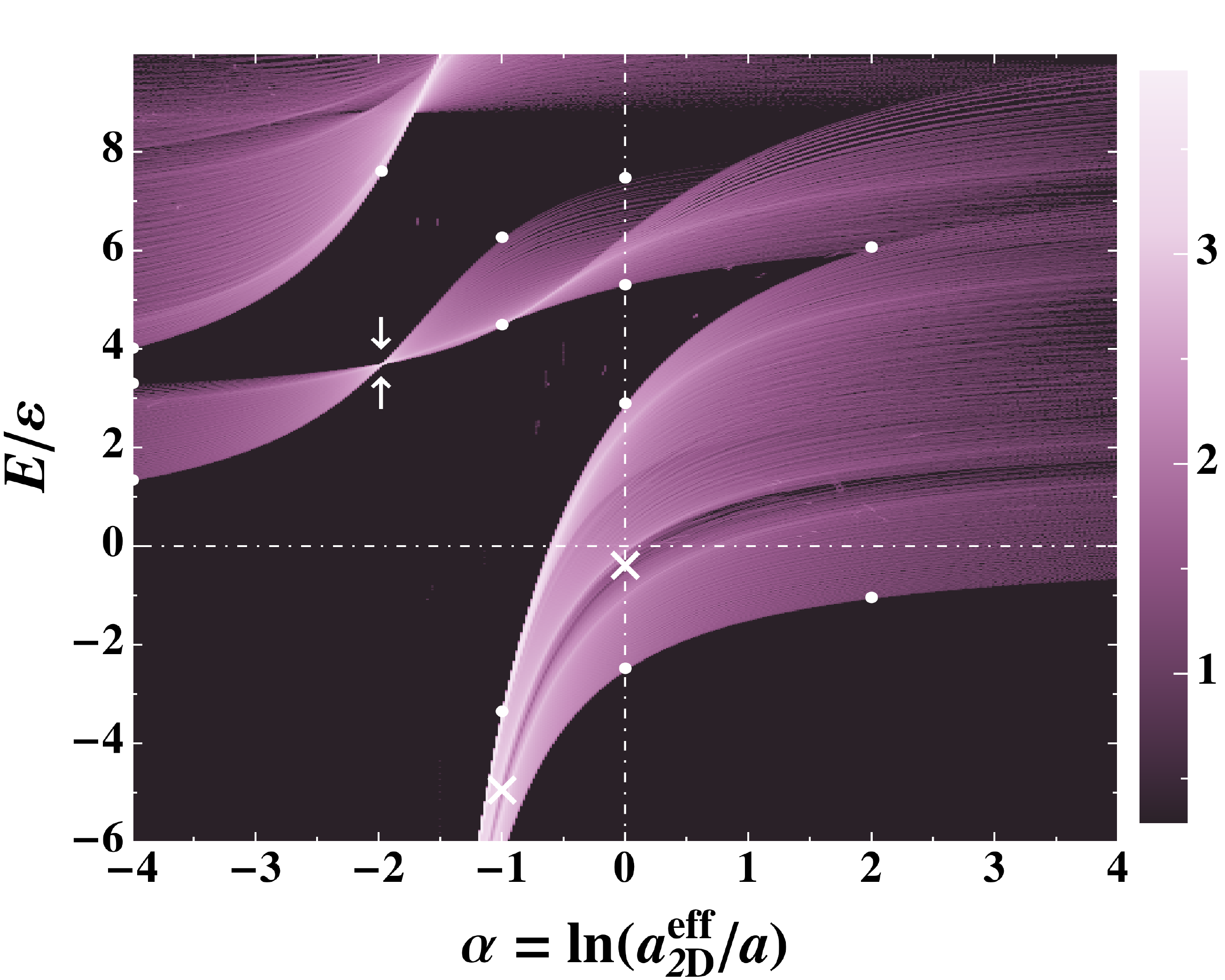}
\caption{(Color online)
Finite-sized kagom\'e lattice: DOS per scatterer in the plane
$[\alpha,E/\varepsilon]$ for a system of $N\!=\!2083$ scatterers
arranged in a kagom\'e lattice inside a disk of radius $R\!=\!24a$.
Results have been obtained with the same method described in Fig.~\ref{FigFiniteSizeSquare}, discretizing the energies with a step of $0.005\varepsilon$ ($\varepsilon\!=\!\hbar^2/ma^2$).
Quasi-Bloch bulk states have been selected choosing $\Gamma_{\rm max}\!\simeq\!0.5\varepsilon/\hbar$ (i.e. $\Gamma_{\rm max}\!\simeq\!1460$Hz for a matter wave of $^{87}$Rb atoms for $a\!=\!500{\rm nm}$).
The color-map is applied to the quantity
$\log_{10}(\frac{N_{\rm p}}{N}\frac{\varepsilon}{\delta\alpha\,\delta E})$,
where $N_{\rm p}$ is the number of selected poles of $G$ within a
rectangular bin of area $\delta\alpha\,\delta E$
($\delta\alpha\!=\!0.02$ and $\delta E\!=\!0.025\varepsilon$).
$\times$, $\bullet$, and arrows indicate, respectively, the positions
of Dirac cones, gap boundaries, and isolated flat band as expected from the
analysis of an infinite system (Figs.~\ref{FigSpecterComparedKagome}).}
\label{FigFiniteSizeKagome}
\end{figure}

The persistence of the spectral features in systems of experimental size have been also investigated.
The density of solutions of Eqs.~\eqref{ConditionFiniteNegative}-\eqref{ConditionFinitePositive} for a system with $\sim2000$ $B$ scatterers is shown in Fig.~\ref{FigFiniteSizeKagome}. The agreement with the predictions for an infinite system is extremely good, demonstrating once again the robustness of our results against finite-size effects.

\section{Conclusions}\label{SecConclusions}

In this work we presented a model for the realization of 2D arbitrary atomic artificial crystals based on the use of two independently trapped atomic species.
This system revealed itself promising as a quantum simulator of bi-dimensional condensed-matter systems.
We discussed how the interplay between scattering length and trappings allows to widely tune the inter-species interaction: a key parameter trough which the system features can be manipulated.
A general theory for finite and infinite periodic systems has been presented, specifying to some illustrative examples for both Bravais and non-Bravais lattices.
We proved the emergence of single and multiple gaps, together with the eventual presence of isolated non-dispersive flat bands. Furthermore we pointed out the existence of Dirac points in atomic artificial graphene and kagom\'e lattices. The robustness of our results for experimentally realizable systems has been tested against both finite size and disorder.
The adaptability of our model makes it suitable for a number of future developments.
New perspectives can be opened considering also the $p$-wave $A\!-\!B$ interaction \cite{NishidaPRL08PRA10} or the $A\!-\!A$ one. The latter, together with the occurrence of flat bands, would make the atomic artificial lattices convenient simulators of 2D strongly correlated systems in which non-trivial effects due to geometrical frustration may arise.

\appendix
\section{Real-to-reciprocal space change}\label{AppChangeOfSpace}

This appendix is devoted to the transformation of sums over real-lattice vectors, appearing in Eqs.\eqref{ConditionBravaisReal}-\eqref{DefinitionTOffDiagReal}-\eqref{DefinitionTDiagReal}, in sums over reciprocal-lattice ones.

\subsection{Bravais terms}\label{AppChangeOfSpaceDiag}

We manipulate here the LHS of Eq.~\eqref{ConditionBravaisReal} or, equivalently, a diagonal element of the matrix $\TT$ defined in Eq.~\eqref{DefinitionTDiagReal}.
In particular we focus on the sum, rewriting $g_0(\RR)$, introduced in Eq.~\eqref{Defg0}, as
\begin{equation}
g_0(\RR) =  \int  \frac{d^2\pp}{(2\pi)^2}\,  \widetilde{g}_0(\pp)\,  e^{i\pp\cdot\RR}
\end{equation}
in terms of its Fourier transform
\begin{equation}
\widetilde{g}_0(\pp)=
\frac{2m}{\hbar^2}\left(\PV\frac{1}{k^2-p^2}-i\frac{\pi}{2k}\delta(k-p)\right),
\end{equation}
where $\PV$ denotes Cauchy's principal value. Once sum and integral are exchanged we obtain
\begin{equation}
\sum_{\RR\in L^*}\!\! g_0(\RR)\, e^{i\qq\cdot\RR}=\!\!
\int\!\!  \frac{d^2\pp}{(2\pi)^2}\, \widetilde{g}_0(\pp)
\left(\sum_{\RR\in L} e^{i(\pp+\qq)\cdot\RR}-1\right),
\end{equation}
where, by adding and subtracting the $\RR\!=\!0$ term, we have now a sum running over the entire $L$.
For such a summation Poisson's identity holds, stating that
\begin{equation}\label{PoissonIdentity}
\sum_{\RR\in L} F(\RR) = \frac{1}{\AAA} \sum_{\KK\in RL} \widetilde{F}(\KK),
\end{equation}
being $\AAA$ the unit cell area in real space. In the case of interest $\widetilde{F}(\KK)\!=\!(2\pi)^2\delta^{(2)}(\pp+\qq-\KK)$ is the Fourier transform of $F(\RR)\!=\!\exp[i(\pp+\qq)\!\cdot\!\RR]$. By integrating out terms involving $\delta$-functions we obtain
\begin{align}
\sum_{\RR\in L^*}\!\! \frac{\pi\hbar^2}{m}\,g_0(\RR)\,& e^{i\qq\cdot\RR}=\,
\frac{2\pi}{\AAA} \sum_{\KK\in RL} \frac{1}{k^2-|\KK-\qq|^2}
\nonumber\\
&+i\frac{\pi}{2}-\PV\!\! \int \frac{d^2\pp}{2\pi}\, \frac{1}{k^2-p^2}.
\end{align}
The principal-valued integral remains to be evaluated.
For an arbitrary choice of $\rho$ such that $0\!<\!\rho\!<\!k$, we have
\begin{equation}
\PV\!\! \int \frac{d^2\pp}{2\pi}\, \frac{1}{k^2-p^2}=
\PV\!\! \int_{p>\rho} \frac{d^2\pp}{2\pi}\, \frac{1}{k^2-p^2}
\,-\frac{1}{2}\ln\!\left(1-\frac{\rho^2}{k^2}\right),
\end{equation}
and the Bravais-element $\TT_{nn}$ becomes
\begin{align}\label{TnnStep1}
\TT_{nn}&=
\ln\left(\frac{e^\gamma}{2}\,k\atwo\right) + \frac{1}{2}\ln\!\left(1-\frac{\rho^2}{k^2}\right)
+ \frac{2\pi}{\AAA}\,\frac{1}{k^2-q^2}
\nonumber\\
& + \frac{2\pi}{\AAA} \sum_{\KK\in RL^*} \frac{1}{k^2-|\KK-\qq|^2}
-\PV\!\! \int_{p>\rho} \frac{d^2\pp}{2\pi}\, \frac{1}{k^2-p^2}.
\end{align}
We need now to introduce two auxiliary quantities:
\begin{align}\label{DefinitionSC}
\mathcal{S}_{\rho,\rm uv}=
\frac{2\pi}{\AAA} \sum_{\KK\in RL^*\setminus \rm uv}& \frac{1}{k^2-|\KK-\qq|^2}
\nonumber\\
&-\PV\! \int_{p>\rho\setminus \rm uv} \frac{d^2\pp}{2\pi}\, \frac{1}{k^2-p^2}
\end{align}
and
\begin{align}\label{DefinitionSU}
S_{\rho,\rm uv}=
\frac{2\pi}{\AAA} \sum_{\KK\in RL^*\setminus \rm uv} &\frac{1}{K^2}
-\int_{p>\rho\setminus \rm uv} \frac{d^2\pp}{2\pi}\, \frac{1}{p^2},
\end{align}
where an arbitrary ultra-violet cut-off uv is added in the domains of sum and integration.
Notice that the second line of Eq.~\eqref{TnnStep1} is exactly $\mathcal{S}_{\rho,\infty}$, that is $\mathcal{S}_{\rho,\rm uv}$ in the limit of a cut-off boundary pushed to infinity.
From definitions~\eqref{DefinitionSC} and~\eqref{DefinitionSU} it follows that
\begin{align}
\mathcal{S}_{\rho,\rm uv}\,+\,&S_{\rho,\rm uv} \xrightarrow[\rm uv\to\infty]{}
- \frac{1}{2}\ln\!\left(\frac{k^2}{\rho^2}-1\right)
\nonumber\\
&+ \frac{2\pi}{\AAA} \sum_{\KK\in RL^*} \left(\frac{1}{k^2-|\KK-\qq|^2}+\frac{1}{K^2}\right).
\end{align}
We thus add and subtract $S_{\rho,\infty}$ to Eq.~\eqref{TnnStep1}
and after some algebraic manipulation we obtain
\begin{align}\label{TnnStep2}
\TT_{nn}&=
\ln(\rho a)-S_{\rho,\infty}+\ln\left(\frac{e^\gamma}{2}\right)
+ \frac{2\pi}{\AAA}\,\frac{1}{k^2-q^2}
\nonumber\\
& + \frac{2\pi}{\AAA} \sum_{\KK\in RL^*} \left(\frac{1}{k^2-|\KK-\qq|^2}+\frac{1}{K^2}\right)+\alpha,
\end{align}
where we introduced the parameter $\alpha\!=\!\ln(\atwo/a)$ and the arbitrary unit of length $a$.
From Eq.~\eqref{TnnStep2} finally follows the definition of
\begin{equation}\label{DefCinfty}
C_\infty=\lim_{\rm uv\to\infty} \ln(\rho a) - S_{\rho,\rm uv} ,
\end{equation}
which numerically converges to a $\rho$-independent quantity determined only by the geometrical properties of the Bravais lattice.

\subsection{Non-Bravais terms}\label{AppChangeOfSpaceOffDiag}
The transformation of the off-diagonal element of $\TT$ introduced in Eq.\eqref{DefinitionTOffDiagReal} follows straightforwardly from the previous one. We write again $g_0(\RR)$ in terms of its Fourier transform $\widetilde{g}_0(\pp)$, obtaining
\begin{align}\label{TnmStep}
\sum_{\RR\in L} & g_0(\RR+\ttt_{nm})\, e^{i\qq\cdot\RR}
\nonumber\\
&=\int \frac{d^2\pp}{(2\pi)^2} \, \widetilde{g}_0(\pp)\, e^{i\pp\cdot\ttt_{nm}}
\sum_{\RR\in L} e^{i(\pp+\qq)\cdot\RR}
\nonumber\\
&=\frac{1}{\AAA} \sum_{\KK\in RL} \widetilde{g}_0(\KK-\qq)\, e^{i(\KK-\qq)\cdot\ttt_{nm}},
\end{align}
where, in the last step, we made use of Poisson's identity as introduced in Eq.~\eqref{PoissonIdentity}.
Finally Eq.~\eqref{DefinitionTOffDiagReciprocal} directly follows from Eq.~\eqref{TnmStep} by writing explicitly $\widetilde{g}_0(\KK-\qq)$.


\begin{thebibliography}{99}

\bibitem{BlochNature12}I. Bloch, J. Dalibard, and S. Nascimb\`ene, Nature Physics {\bf 8}, 267 (2012).

\bibitem{ChinRMP10}C. Chin, R. Grimm, P. S. Julienne, and E. Tiesinga, Rev. Mod. Phys. {\bf 82}, 1225 (2010).

\bibitem{CitDIR}M. Marinescu and L. You, Phys. Rev. Lett. {\bf 81}, 4596 (1998);
Z. Y. Shi, R. Qi, and H. Zhai, Phys. Rev. A {\bf 85}, 020702(R) (2012);
N. Bartolo, D. J. Papoular, L. Barbiero, C. Menotti, and A. Recati, Phys. Rev. A {\bf 88}, 023603 (2013).

\bibitem{OlshaniiPRL98}M. Olshanii, Phys. Rev. Lett. {\bf 81}, 938 (1998).

\bibitem{CitQPT}M. P. A. Fisher, P. B. Weichman, G. Grinstein, and D. S. Fisher, Phys. Rev. B {\bf 40}, 546 (1989);
M. Greiner, O. Mandel, T. Esslinger, T. W. H\"ansch, and I. Bloch, Nature {\bf 415}, 40 (2002).

\bibitem{NovoselovScience04}K. S. Novoselov, A. K. Geim, S. V. Morozov, D. Jiang, Y. Zhang, S. V. Dubonos, I. V. Gregorieva, and A. A. Firsov, Science \textbf{306}, 666 (2004).

\bibitem{GeimNature07}A. K. Geim and K. S. Novoselov, Nature materials {\bf 6}, 183 (2007).

\bibitem{CastroNetoRMP09}A. H. Castro Neto, F. Guinea, N. M. R. Peres, K. S. Novoselov, and A. K. Geim, Rev. Mod. Phys. {\bf 81}, 109 (2009).

\bibitem{PoliniNature13}M. Polini, F. Guinea, M. Lewenstein, H. C. Manoharan, and V. Pellegrini, Nature nanotechnology {\bf 8}, 625 (2013).

\bibitem{LamporesiPRL10}G. Lamporesi, J. Catani, G. Barontini, Y. Nishida, M. Inguscio, and F. Minardi, Phys. Rev. Lett. {\bf 104}, 153202 (2010).

\bibitem{GadwayPRL11}B. Gadway, D. Pertot, J. Reeves, M. Vogt, and D. Schneble, Phys. Rev. Lett. {\bf 107}, 145306 (2011)

\bibitem{BakrScience10}W. S. Bakr, A. Peng, M. E. Tai, R. Ma, J. Simon, J. I. Gillen, S. F\"olling, L. Pollet, and M. Greiner, Science {\bf 329}, 547 (2010).

\bibitem{NogrettePRX14}F. Nogrette, H. Labuhn, S. Ravets, D. Barredo, L. B\'eguin, A. Vernier, T. Lahaye, and A. Browaeys, Phys. Rev. X {\bf 4}, 021034 (2014).

\bibitem{GavishPRL05}U. Gavish and Y. Castin, Phys. Rev. Lett. {\bf 95}, 020401 (2005).

\bibitem{AntezzaPRA10}M. Antezza, Y. Castin, and D. A. W. Hutchinson, Phys. Rev. A \textbf{82}, 043602 (2010).
We point out a misprint in Eq.~(47) where a factor 8 is missing in front of $\delta^2$.

\bibitem{BartoloEPL14}N. Bartolo and M. Antezza, Europhys. Lett. {\bf 107}, 30006 (2014).

\bibitem{PetrovPRA01}D. S. Petrov and G. V. Shlyapnikov, Phys. Rev. A {\bf 64}, 012706 (2001).

\bibitem{MassignanPRA06}P. Massignan and Y. Castin, Phys. Rev. A {\bf 74}, 013616 (2006).

\bibitem{NishidaPRL08PRA10}Y. Nishida and S. Tan, Phys. Rev. Lett. {\bf 101}, 170401 (2008); Phys. Rev. A {\bf 82}, 062713 (2010).

\bibitem{BartoloPreparation}N. Bartolo and M. Antezza, in preparation.

\bibitem{OlshaniiPRL01JPB07}M. Olshanii and L. Pricoupenko, Phys. Rev. Lett. {\bf 88}, 010402 (2001);
L. Pricoupenko and M. Olshanii, J. Phys. B {\bf 40}, 2065 (2007).

\bibitem{KittelBook}C. Kittel, \textit{Introduction to Solid State Physics} (John Wiley \& Sons Inc, 2004).

\bibitem{CarusottoPRA08AntezzaPRL09}I. Carusotto, M. Antezza, F. Bariani, S. De Liberato, and C. Ciuti, Phys. Rev. A \textbf{77}, 063621 (2008);
M. Antezza and Y. Castin, Phys. Rev. Lett. \textbf{103}, 123903 (2009).

\bibitem{AntezzaPRA09}M. Antezza and Y. Castin, Phys. Rev. A \textbf{80}, 013816 (2009).

\bibitem{AntezzaPRA13}M. Antezza and Y. Castin, Phys. Rev. A {\bf 88}, 033844 (2013).

\bibitem{ZhuPRL07}S.-L. Zhu, B. Wang, and L.-M. Duan, Phys. Rev. Lett. {\bf 98}, 260402 (2007).

\bibitem{TarruellNature12}L. Tarruell, D. Greif, T. Uehlinger, G. Jotzu, and T. Esslinger, Nature {\bf 483}, 302 (2012).

\bibitem{WuPRL07PRB08}C. Wu, D. Bergman, L. Balents, and S. Das Sarma, Phys. Rev. Lett. \textbf{99}, 070401 (2007);
C. Wu and S. Das Sarma, Phys. Rev. B \textbf{77}, 235107 (2008).

\bibitem{JacqminPRL14}T. Jacqmin, I. Carusotto, I. Sagnes, M. Abbarchi, D. D. Solnyshkov, G. Malpuech, E. Galopin, A. Lema\^itre, J. Bloch, and A. Amo, Phys. Rev. Lett. {\bf 112}, 116402 (2014).

\bibitem{CitMotionDirac}G. Montambaux, F. Pi\'echon, J.-N. Fuchs, and M. O. Goerbig, Eur. Phys. J. B {\bf 72}, 509 (2009);
L.-K. Lim, J.-N. Fuchs, and G. Montambaux, Phys. Rev. Lett. {\bf 108}, 175303 (2012);
J. Iba\~nez-Azpiroz, A. Eiguren, A. Bergara, G. Pettini, and M. Modugno, Phys. Rev. A {\bf 88}, 033631 (2013).

\bibitem{CitFrustration}L. Santos, M. A. Baranov, J. I. Cirac, H.-U. Everts, H. Fehrmann, and M. Lewenstein, Phys. Rev. Lett. {\bf 93}, 030601 (2004);
G.-B. Jo, J. Guzman, C. K. Thomas, P. Hosur, A. Vishwanath, and D. M. Stamper-Kurn, Phys. Rev. Lett. {\bf 108}, 045305 (2012).

\end{thebibliography}
\end{document}